\documentclass[preprint,3p,12pt,sort&compress]{elsarticle}
\usepackage{mathrsfs}
\usepackage{amsmath}
\usepackage{stmaryrd}
\usepackage{bbding}
\usepackage{dcolumn}
\usepackage{graphicx}
\usepackage{amsfonts}
\usepackage{amssymb}
\usepackage{psfrag}
\usepackage{wrapfig}
\usepackage{subfigure}
\usepackage{makeidx}
\usepackage{bm}
\usepackage{epsf}
\usepackage{epsfig}
\usepackage{setspace}
\usepackage{graphicx}
\usepackage{epstopdf}
\usepackage{psfrag}
\usepackage{subfigure}
\usepackage{color}
\usepackage{enumerate}
\usepackage{verbatim} 
\usepackage{booktabs}
\usepackage[utf8]{inputenc}
\usepackage{amsthm}
\usepackage{url}

\newtheorem{theorem}{Theorem}

\theoremstyle{remark}

\definecolor{orange}{rgb}{1,0.5,0}

\begin{document}

\title{A flux reconstruction stochastic Galerkin scheme for hyperbolic conservation laws}

\author[KIT]{Tianbai Xiao\corref{cor}}
\ead{tianbaixiao@gmail.com}

\author[KIT]{Jonas Kusch}

\author[Leuven]{Julian Koellermeier}

\author[KIT]{Martin Frank}

\cortext[cor]{Corresponding author}
\address[KIT]{Karlsruhe Institute of Technology, Karlsruhe, Germany}
\address[Leuven]{KU Leuven, Leuven, Belgium}

\begin{abstract}

The study of uncertainty propagation poses a great challenge to design numerical solvers with high fidelity.
Based on the stochastic Galerkin formulation, this paper addresses the idea and implementation of the first flux reconstruction scheme for hyperbolic conservation laws with random inputs.
Unlike the finite volume method, the treatments in physical and random space are 
consistent, e.g., the modal representation of solutions based on an orthogonal polynomial basis and the nodal representation based on solution collocation points.
Therefore, the numerical behaviors of the scheme in the phase space can be designed and understood uniformly.
A family of filters is extended to multi-dimensional cases to mitigate the well-known Gibbs phenomenon arising from discontinuities in both physical and random space.
The filter function is switched on and off by the dynamic detection of discontinuous solutions, and a slope limiter is employed to preserve the positivity of physically realizable solutions.
As a result, the proposed method is able to capture stochastic cross-scale flow evolution where resolved and unresolved regions coexist.
Numerical experiments including wave propagation, Burgers' shock, one-dimensional Riemann problem, and two-dimensional shock-vortex interaction problem are presented to validate the scheme.
The order of convergence of the current scheme is identified.
The capability of the scheme for simulating smooth and discontinuous stochastic flow dynamics is demonstrated.
The open-source codes to reproduce the numerical results are available under the MIT license \cite{csmm2021}.

\end{abstract}


\maketitle

\section{Introduction}

The thriving discipline of uncertainty quantification (UQ) has contributed to applications in meteorology, particle physics, chemistry, bioinformatics, etc \cite{smith2013uncertainty}.
In this paper, we focus on the propagation of randomness in stochastic conservation laws, i.e.,
\begin{equation}
\begin{aligned}
    & \partial_{t} \mathbf{u}(t, \mathbf{x}, \mathbf z)+\nabla \cdot \mathbf{f}(\mathbf{u}(t, \mathbf{x}, \mathbf z))=\mathbf{0}, \quad t \in (0,T], \ \mathbf x \in \boldsymbol \Omega, \ \mathbf z \in \boldsymbol \Upsilon, \\
    & \mathcal{B}(\mathbf u)=0, \quad t \in (0,T], \ \mathbf x \in \partial \boldsymbol \Omega, \ \mathbf z \in \boldsymbol \Upsilon, \\
    & \mathbf u(0, \mathbf{x}, \mathbf z) = \mathbf u_0, \quad \mathbf x \in \boldsymbol \Omega, \ \mathbf z \in \boldsymbol \Upsilon,
\end{aligned}
\label{eqn:conservation laws}
\end{equation}
where $T\in \mathbb R^+$ is the evolution time, $\boldsymbol \Omega \subset \mathbb{R}^{d}$ is the physical space of dimension $d$, $\boldsymbol \Upsilon \subset \mathbb{R}^{l}$ is the random space of dimension $l$, and $\mathcal B$ is the boundary operator.

Depending on the treatment of the random variable, the methods for uncertainty quantification can be classified into intrusive and non-intrusive ones.
A typical non-intrusive strategy is Monte Carlo sampling.
Based on a prescribed probability distribution, many realizations of random inputs are produced.
The deterministic computation is performed in each realization and a post-processing is conducted afterwards to estimate uncertainties.
The Monte Carlo methods are intuitive and easy to implement, but a large number of realizations is needed due to the slow convergence with respect to the sampling size.
This remains true for other variants like quasi or multi-level Monte-Carlo, which differ in the nodes and weights that are used in the post-processing \cite{giles2015multilevel}.

On the other hand, intrusive methods work in a way such that the original governing equation system in Eq.(\ref{eqn:conservation laws}) is reformulated.
One commonly used intrusive strategy is the stochastic Galerkin (SG) method, in which the stochastic solutions are expressed into generalized polynomial chaos (gPC) of the input random parameters \cite{xiu2010numerical}. 
The solution of Eq.(\ref{eqn:conservation laws}) is thus transformed into the solution of coefficients in the polynomial expansions.
As the residual of the governing equations is orthogonal to the linear space spanned by the polynomial chaos, the spectral convergence can be achieved provided that the solution depends smoothly on the random parameters.

The stochastic Galerkin method has been successfully applied to solve hyperbolic equations \cite{knio2001stochastic,pettersson2014stochastic,jin2015asymptotic,wu2017stochastic,schlachter2018hyperbolicity,hu2019stochastic,kusch2020filtered,xiao2021plasma}.
It is noticeable that these methods employ finite difference or finite volume methods to discretize the balance laws of the gPC coefficients.
Although this has proven to be an effective strategy, the different discretization strategies, i.e., the finite difference in physical space and the spectral representation in random space, make it indirect to understand the behavior of the numerical scheme consistently.
Besides, most of the methods above hold no more than second order of accuracy.
It has been noted in \cite{pettersson2009numerical,offner2018stability} that the spatial discretization has significant effects on the solution quality in random space. 
The diffusive behavior of low-order methods can heavily smear out the solution.
While it is possible to develop higher-order methods based on the finite difference or finite volume framework, the lack of ability to handle irregular geometry in the finite difference method and the non-compact stencils used in the traditional finite volume method prevent such extensions from being universally applicable.

The inherited high resolution and low dissipation of higher-order computational methods enable high-fidelity simulation of intricate flows in turbulence, acoustics, magnetohydrodynamics, etc \cite{barth2013high}.
It motivates a unified spectral discretization for the stochastic Galerkin system, which leads to compatible accuracy in stochastic and spatial domain.
This has been realized in \cite{xiu2003modeling,sousedik2016stochastic} for stochastic Navier-Stokes equations.
However, in a hyperbolic system, discontinuous solutions can emerge from a smooth initial field, 
and the well-known Gibbs phenomenon brings tremendous challenges for spectral methods to capture the discontinuities in both physical and random space. 
To the best of the authors' knowledge, only two research groups have addressed this issue following the discontinuous Galerkin (DG) approach \cite{cockburn2012discontinuous,hesthaven2007nodal}.
D\text{ü}rrw\text{ä}chter et. al. \cite{durrwachter2020hyperbolicity,durrwachter2021high} developed a discontinuous stochastic Galerkin method for stochastic fluid dynamic equations.
Donoghue and Yano \cite{donoghue2021spatio} proposed a similar methodology, while the focus is on the adaptive refinement of spatial mesh or polynomial chaos to control the numerical error.


The success of DG method is attributed to the unified consideration of the spatial discretization and the spectral decomposition.
Within each element, the solutions are approximated by polynomials and are allowed to be discontinuous across cell boundaries, which encourages the capturing of sharp structures that arise in hyperbolic systems.
Based on similar views, the flux reconstruction (FR) approach proposed by Huynh et al. \cite{huynh2007flux,vincent2011new,castonguay2013energy} provides profound insight into constructing high-order methods for transport equations.
It establishes a universal framework, where several existing approaches, including the nodal DG and the spectral difference (SD) \cite{kopriva1996conservative,liu2006spectral} methods, can be cast within by choosing different correction fields of Lagrange polynomials.
The intrinsic connections between FR and DG or SD methods have been analyzed in \cite{de2014connections,cox2021accuracy}.

It is desirable to design the solution algorithm that equips consistent accuracy in time, space, and random domain for stochastic conservation laws.
In this paper, we employ the flux reconstruction methodology as building blocks and develop the high-order stochastic Galerkin method for hyperbolic conservation laws.
A family of multi-dimensional filters is developed to mitigate the oscillating solutions around discontinuities in the physical-random space.
The filter function is dynamically dispatched based on a detector of discontinuous solutions to optimize the numerical dissipation.
A slope limiter is applied to the nodal solutions to ensure the positivity of physically realizable solutions (e.g., the density and temperature in the Euler equations) and preserve the hyperbolicity of the stochastic Galerkin system.
The proposed algorithm can be understood uniformly as a spectral method within modal expansions and as a collocation method upon nodal solution points.
The discontinuity capturing strategy is naturally incorporated into the solution algorithm based on the nodal-modal transformation.
As a result, the proposed method is able to capture the cross-scale stochastic dynamics where resolved and unresolved regions coexist inside a flow field.

The rest of the paper is structured as follows.
Section \ref{sec:stochastic galerkin} introduces the generalized polynomial chaos and stochastic Galerkin formulation of hyperbolic conservation laws.
Section \ref{sec:flux reconstruction} presents the implementation of the flux reconstruction framework.
Section \ref{sec:shock capturing} expounds the strategy for capturing discontinuous solutions using filters.
Section \ref{sec:numerical experiments} includes the numerical experiments to demonstrate the performance of the new scheme.
The paper ends with a short conclusion.
The source codes to produce the numerical results are hosted on GitHub and distributed under the MIT license \cite{csmm2021}.

\section{Stochastic Galerkin Method}\label{sec:stochastic galerkin}

\subsection{Formulation}

The stochastic Galerkin method employs the generalized polynomial chaos (gPC) to describe the evolution of stochastic solutions.
A spectral representation of degree $N$ is introduced in the random space as
\begin{equation}
\mathbf u(t,\mathbf x,\mathbf z) \simeq \mathbf u^{N}(t,\mathbf x,\mathbf z) = \sum_{|k|=0}^N \hat {\mathbf u}_{k} (t,\mathbf x) \Phi_k (\mathbf z) = \hat{\boldsymbol v}(t,\mathbf x) \boldsymbol \Phi(\mathbf z) ,
\label{eqn:polynomial chaos}
\end{equation}
where $\hat{\mathbf u}_k=(\hat u_{1k}, \hat u_{2k}, \cdots, \hat u_{Sk})^T$ are the expansion coefficients of conservative variables in the polynomial chaos, and they are also called moments of the stochastic Galerkin system.
The number of states in the solution vector is equal to $S$, which takes unit value for scalar conservation laws.
The index $k$ can be a scalar or a $P$-dimensional vector $k = (k_1,k_2,\cdots,k_P )^T$ with $|k| = k_1 + k_2 + \cdots + k_P$.
The matrix $\hat{\boldsymbol v}= \{\hat{\mathbf u}_k, | k| \leq N \}$ denotes a collection of the gPC coefficients at all orders.

The orthogonal polynomial basis $\boldsymbol \Phi$ satisfies the following constraints,
\begin{equation}
\mathbb{E}[\Phi_ j (\mathbf z) \Phi_ k (\mathbf z)] = \delta_{ j  k}, \quad 0 \leq | j|, | k| \leq N.
\end{equation}
The expected value defines a scalar product,
\begin{equation}
\mathbb{E}[\Phi_ j (\mathbf z) \Phi_ k (\mathbf z)] = \int_{\boldsymbol\Upsilon} \Phi_{ j}(\mathbf z) \Phi_{ k}(\mathbf z) \varrho(\mathbf z) d \mathbf z,
\label{eqn:continuous expectation value}
\end{equation}
where $\varrho(\mathbf z): \boldsymbol\Upsilon \rightarrow[0, \infty) $ is the probability density function.
In practice, the above integral can be evaluated analytically or with the help of a numerical quadrature rule, i.e.,
\begin{equation}
\mathbb{E}[\Phi_ j (\mathbf z) \Phi_ k (\mathbf z)] = \sum_{q=1}^{N_q} \Phi_{ j}(\mathbf z_q) \Phi_{ k}(\mathbf z_q) w(\mathbf z_q) ,
\label{eqn:discrete expectation value}
\end{equation}
where $w(\mathbf z_q)$ is the corresponding quadrature weight function in random space.
In the following we adopt a uniform notation $\mathbb{E}[\Phi_ j (\mathbf z) \Phi_ k (\mathbf z)]=\langle \Phi_ j \Phi_ k \rangle$ to denote the integrals over random space from Eq.(\ref{eqn:continuous expectation value}) and (\ref{eqn:discrete expectation value}).

Plugging Eq.(\ref{eqn:polynomial chaos}) into Eq.(\ref{eqn:conservation laws}) and projecting the resulting residual
to zero, we get the conservation laws in the stochastic Galerkin formulation,
\begin{equation}
\begin{aligned}
& \partial_{t} \hat{\mathbf{u}}_{k} + \langle \nabla \cdot \mathbf{f}\left(\mathbf{u}^{N}\right) \Phi_{k} \rangle =\mathbf{0}, \quad t \in (0,T], \ \mathbf x \in \boldsymbol \Omega, \\
& \langle \mathcal B(\mathbf u_N) \Phi_k \rangle = 0, \quad t \in (0,T], \ \mathbf x \in \partial \boldsymbol \Omega, \\
& \hat{\mathbf{u}}_{k}(t=0, \mathbf{x})=\langle\mathbf u_0(\mathbf x, \mathbf z) \Phi_k \rangle, \quad \mathbf x \in \boldsymbol \Omega.
\end{aligned}
\label{eqn:sg conservation laws}
\end{equation}
The stochastic Galerkin approach provides a desirable accuracy for the smooth solution in random space, where the residual of the governing equations is orthogonal to the linear space spanned by the gPC polynomials \cite{xiu2010numerical}.

\subsection{Challenge}

While the stochastic Galerkin method has been successfully applied to various settings, its application in hyperbolic problems faces two main challenges:
First, the SG system for the gPC coefficients in Eq.(\ref{eqn:sg conservation laws}) is not necessarily hyperbolic, leading to a possible breakdown of the numerical method \cite{abgrall2017uncertainty,Poette2009}. 
Strategies to preserve hyperbolicity of the SG system include the intrusive polynomial moment (IPM) method \cite{Poette2009}, the Roe transformation method \cite{pettersson2014stochastic}, and the hyperbolicity-preserving limiter \cite{schlachter2018hyperbolicity}.
The IPM method is a generalization of stochastic Galerkin, which performs the gPC expansion on the entropy variables instead of the original conservative variables. 
Similarly, the Roe transformation method performs the expansion on the Roe variables.
The hyperbolicity-preserving SG method employs a bound-preserving limiter to enforce positive moments of thermodynamic variables, which in turn guarantee the hyperbolicity of the SG system.

The second challenge is that the modal approximation suffers from the Gibbs phenomenon when the solution exhibits sharp gradients \cite{le2004uncertainty}.
Strategies to mitigate spurious artifacts from the Gibbs phenomenon in the random space have recently been developed. 
The multi-element SG method \cite{Wan2006,tryoen2010intrusive} utilizes $h$-refinement in the stochastic space, which is less prone to oscillations.
The filtered SG and IPM methods are proposed in \cite{kusch2020filtered,alldredge2021realizable}, where a filtering step is applied to the solution in between time steps. 
In addition, stochastic adaptivity \cite{tryoen2012adaptive,buerger2014hybrid,meyer2020posteriori,kusch2020intrusive} can be employed to increase the truncation order in oscillatory regions. For the IPM method, certain choices of the entropy mitigate oscillations \cite{kusch2019maximum}.
It is a natural idea to combine different strategies for a better control of the numerical accuracy.
As an example, in \cite{kusch2021oscillation} the multi-element approach is extended to IPM and a filter step is performed after applying the bound-preserving limiter, which reduces the oscillations while maintaining hyperbolicity. A strategy of picking a sufficiently strong filter strength to preserve physical bounds of the solution is proposed in \cite{xiao2021stochastic}.

The physical realizablity (e.g. the positivity of certain thermodynamic variables) and robustness of solutions are closely coupled.
The Gibbs phenomenon can lead to unrealizable solutions and thus break up the hyperbolicity.
It is desirable to consider the hyperbolicity preservation and mitigation of the Gibbs phenomenon uniformly in the solution algorithm.
In this paper, we will develop multi-dimensional filters that can mitigate spurious artifacts from the Gibbs phenomenon in both physical and random space.
A multi-dimensional slope limiter is applied simultaneously to ensure the realizability of physical solutions.
The detailed strategy will be illustrated in section \ref{sec:shock capturing}.

\section{Flux Reconstruction Framework}\label{sec:flux reconstruction}


\subsection{Formulation}

Considering $N_i$ non-overlapping cells in the domain $\boldsymbol{\Omega}=\bigcup_{i=1}^{N_i} \boldsymbol{\Omega}_{i}$,
%
we approximate the solution of the conservation laws with piecewise polynomials, i.e., 
\begin{equation}
    \mathbf u \approx \bigoplus_{i=1}^{N_i} \hat {\boldsymbol v}_{i} \mathbf \Phi , \quad \mathbf f \approx \bigoplus_{i=1}^{N_i} \hat {\boldsymbol f}_{i} \mathbf \Phi.
\end{equation}

For convenience, the standard element in the reference space can be introduced based on the transformation of coordinates,
\begin{equation}
    \mathbf x_i = \mathbf \Theta_i(\mathbf r) = \sum_{j=1}^{N_v} \lambda_j(\mathbf r) \mathbf x_{i,j},
\end{equation}
where $\{\mathbf x, \mathbf r\}$ represent the global and local coordinates of a point in the element $\mathbf \Omega_i$.
These two coordinates can be connected by the vertex coordinates $\lambda_j$, which are built upon $N_v$ vertices and their global coordinates $\mathbf x_{i,j}$.
For elements of different shapes, the vertex coordinates take different forms, e.g.,
\begin{equation}
    \lambda_1=\frac{1-r}{2}, \ \lambda_2=\frac{1+r}{2},
\end{equation}
in one-dimensional line elements,
\begin{equation}
    \lambda_1=-\frac{r+s}{2},\ \lambda_2=\frac{r+1}{2},\ \lambda_3=\frac{s+1}{2},
\end{equation}
in isosceles right triangle elements where $\mathbf r=(r,s)^T$, and the bi-linear rectangle shape functions,
\begin{equation}
\begin{aligned}
    \lambda_1=\frac{(r-1)(s-1)}{4},\ \lambda_2=\frac{(r+1)(1-s)}{4},\\
    \lambda_3=\frac{(r+1)(s+1)}{4}, \ \lambda_4=\frac{(1-r)(s+1)}{4},
\end{aligned}
\end{equation}
in square elements.

Therefore, the stochastic Galerkin conservation laws in the reference space read
\begin{equation}
    \frac{\partial \hat{\boldsymbol v}^{\delta}}{\partial t}=-\nabla_{\mathbf r} \cdot \hat{\boldsymbol f}^{\delta},
    \label{eqn:fr update}
\end{equation}
where $\hat{\boldsymbol v}^\delta$ denotes the matrix of all the gPC coefficients in the reference space, and $\hat{\boldsymbol f}^\delta$ are the numerical fluxes.

\subsection{Discontinuous flux}\label{sec:discontinuous flux}

In the flux reconstruction method, the solution is approximated by piecewise polynomials in physical space.
For brevity, we consider one-dimensional geometry first to illustrate the solution algorithm.
Defining the Lagrange polynomials based on $N_p$ solution points,
\begin{equation}
    \ell_{j}=\prod_{k=1, k \neq j}^{N_p}\left(\frac{r-r_{k}}{r_{j}-r_{k}}\right),
\end{equation}
the conservative variables in the element $\Omega_i$ can be represented as,
\begin{equation}
    \hat{\boldsymbol v}^{\delta}(t, r)=\sum_{j=1}^{N_p} \hat{\boldsymbol v}^{\delta}(t,r_j) \ell_{j}(r).
\end{equation}
The fluxes at these solution points can then be determined and transformed via
\begin{equation}
    \hat{\boldsymbol f}^{\delta D}(t, r)=\frac{\hat{\boldsymbol f} \left( \hat{\boldsymbol v} \left(t,\mathbf \Theta_{i}(r)\right) \right) }{J_{i}},
    \label{eqn:local direct flux}
\end{equation}
where $\hat{\boldsymbol f}$ is the flux function related to the specific governing equations, and $J_i=(x_{i+1/2}-x_{i-1/2})/2$ is the Jacobian.
Therefore, the flux polynomials can be constructed as,
\begin{equation}
    \hat{\boldsymbol f}^{\delta D}(t, r)=\sum_{j=1}^{N_p} \hat{\boldsymbol f}^{\delta D}(t,r_j) \ell_{j}(r),
    \label{eqn:discontinuou flux polynomial}
\end{equation}
where $\hat{\boldsymbol f}^{\delta D}(t,r_j)$ denotes the evaluated flux calculated by Eq.(\ref{eqn:local direct flux}) at solution point $r_j$ and time $t$.
The notation $\delta D$ implies that such a flux is basically discontinuous since it is derived directly from piecewise discontinuous solutions $\hat {\boldsymbol v}^\delta$.

\subsection{Interactive flux}

It is noticeable that the discontinuous flux polynomials in Eq.(\ref{eqn:local direct flux}) are of the same degree of freedom $N_p$ as solutions, which fail to build the numerical solution with $N_p+1$ order of accuracy.
Besides, the numerical treatment does not take the information from adjacent cells into consideration and can by no means deal with boundary conditions.
A natural idea is to introduce a correction term of order $N_p+1$ to the transformed discontinuous fluxes, i.e.,
\begin{equation}
    \hat{\boldsymbol f}^{\delta} = \hat{\boldsymbol f}^{\delta D} + \hat{\boldsymbol f}^{\delta C}.
\end{equation}
The total fluxes are expected to equal the correct interactive fluxes at cell boundaries, and to preserve similar in-cell profiles of discontinuous fluxes.
A feasible approach, as proposed in \cite{huynh2007flux}, is to introduce two symmetric auxiliary functions $\{ h_L, h_R \}$, which satisfy the following restrictions,
\begin{equation}
\begin{aligned}
    & h_L(r) = h_R(-r), \\
    & h_L(-1)=1, \ h_R(-1)=0, \\
    & h_L(1)=0, \ h_R(1)=1. \\
\end{aligned}
\end{equation}
The corresponding correction flux can then be reconstructed as
\begin{equation}
    \hat{\boldsymbol f}^{\delta C} = (\hat{\boldsymbol f}^{\delta I}_L - \hat{\boldsymbol f}^{\delta D}_L) h_L + (\hat{\boldsymbol f}^{\delta I}_R - \hat{\boldsymbol f}^{\delta D}_R) h_R.
    \label{eqn:interactive flux}
\end{equation}
Here $\{ \hat{\boldsymbol f}^{\delta D}_L,\hat{\boldsymbol f}^{\delta D}_R \}$ are the reconstructed discontinuous fluxes from the Lagrange interpolation at the left and right boundary of the element,
and $\{ \hat{\boldsymbol f}^{\delta I}_L,\hat{\boldsymbol f}^{\delta I}_R \}$ are the interactive fluxes at the boundaries.
Such fluxes can be obtained by nonlinear flux solvers, e.g. the Lax-Friedrichs and Roe's method.

\subsection{Total flux}

Given the total flux $\hat{\boldsymbol f}^\delta$, its derivatives can be expressed as
\begin{equation}
\begin{aligned}
    \frac{\partial \hat{\boldsymbol f}^{\delta}}{\partial r} = \frac{\partial \hat{\boldsymbol f}^{\delta D}}{\partial r} + \frac{\partial \hat{\boldsymbol f}^{\delta C}}{\partial r}. \\
\end{aligned}
\end{equation}
It can be evaluated by calculating the divergences of the Lagrange polynomials and the correction functions at each solution point $r_j$, i.e.
\begin{equation}
    \frac{\partial \hat{\boldsymbol f}^{\delta}}{\partial r}(r_j)=\sum_{k=1}^{N_p} \hat{\boldsymbol f}_{k}^{\delta D} \frac{\mathrm{d} \ell_{k}}{\mathrm{~d} r}\left(r_{j}\right)+\left(\hat{\boldsymbol f}_{L}^{\delta I}-\hat{\boldsymbol f}_{L}^{\delta D}\right) \frac{\mathrm{d} h_{L}}{\mathrm{~d} r}\left(r_{j}\right)+\left(\hat{\boldsymbol f}_{R}^{\delta I}-\hat{\boldsymbol f}_{R}^{\delta D}\right) \frac{\mathrm{d} h_{R}}{\mathrm{~d} r}\left(r_{j}\right).
\end{equation}
Till now, we have completed the construction of the right-hand side of the governing equations.
Appropriate numerical integrators can be chosen to compute the time-marching solutions.

\subsection{Multi-dimensional extension}

The above flux reconstruction procedures can be extended to multi-dimensional cases.
Inside the element $\mathbf \Omega_i$, we approximate the solutions as,
\begin{equation}
    \hat{\boldsymbol v}^{\delta}(t, \mathbf r)=\sum_{j=1}^{N_p} \hat{\boldsymbol v}_{j}^{\delta}(t) \ell_{j}(\mathbf r),
\end{equation}
where $\ell_j(\mathbf r)$ denotes the the multi-dimensional Lagrange polynomials, and $N_p$ is the number of solution points.
If tensorized elements are considered, the above expansion can be simplified as the product of one-dimensional Lagrange polynomials.
For example, in a quadrilateral element, the solution expansion takes the form,
\begin{equation}
    \hat{\boldsymbol v}^{\delta}(t, r, s)=\sum_{j=1}^{\sqrt{N_p}}\sum_{k=1}^{\sqrt{N_p}} \hat{\boldsymbol v}_{j,k}^{\delta}(t) \ell_{j}(r)\ell_{k}(s),
\end{equation}
where $\mathbf r=(r,s)$.
The Lagrange polynomials in a generic element can be evaluated by the nodal-modal transformation with the help of the Vandermonde matrix \cite{hesthaven2007nodal}.
Therefore, the right-hand side of the governing equation in the flux reconstruction formulation becomes
\begin{equation}
\begin{aligned}
    \frac{\partial \hat{\boldsymbol v}^\delta}{\partial t}(\mathbf r_j) =& -\nabla_{\mathbf r} \cdot \hat{\boldsymbol F}^\delta (\mathbf r_j) \\
    =& \sum_{k=1}^{N_p} \hat{\boldsymbol F}_{k}^{\delta D} \cdot \nabla_{\mathbf r} \ell_{k}\left(\mathbf r_{j}\right)
    +\sum_{f=1}^{N_f} \sum_{k=1}^{N_{f p}}\left[\left(\hat{\boldsymbol{F}}_{f, k}^{I}-\hat{\boldsymbol{F}}_{f, k}^{\delta D}\right) \cdot {\mathbf{n}}^\delta_{f, k}\right] \nabla_{\mathbf r} \cdot \mathbf{h}_{f,k}(\mathbf{r}_{j}).
\end{aligned}
\end{equation}
where $N_f$ is the number of faces and $N_{fp}$ is the number of flux points at each face.
The flux tensor takes $\hat{\boldsymbol F}^\delta=(\hat{\boldsymbol f}^\delta,\hat{\boldsymbol g}^\delta)$ in the two-dimensional case and $\hat{\boldsymbol F}^\delta=(\hat{\boldsymbol f}^\delta,\hat{\boldsymbol g}^\delta,\hat{\boldsymbol h}^\delta)$ in the three-dimensional case.
The unit normal vector $\hat{\mathbf{n}}_{f, k}$ points outwards of the element.
The correction function $\mathbf h_{f,k}$ at $k$-th flux point of $f$-th face is a vector, which satisfies the following constraints,
\begin{equation}
    \mathbf{h}_{f, k}\left(\mathbf{r}_{j, l}\right) \cdot {\mathbf{n}}^\delta_{j, l}=\left\{\begin{array}{ll}
    1, & \text { if } f=j \text { and } k=l, \\
    0, & \text { if } f \neq j \text { or } k \neq l.
\end{array}\right.
\end{equation}

\section{Discontinuity Capturing Strategy}\label{sec:shock capturing}

In this section, we present the detailed strategy for capturing discontinuous solutions robustly and maintaining the hyperbolicity of the system.
A series of filters that can be applied in the multi-dimensional physical-random space is introduced to reduce the Gibbs phenomenon.
A detector of discontinuity is employed to adapt numerical dissipation based on local flow conditions and maintain the optimal accuracy.
Besides, a positivity-preserving limiter is built to enforce the realizability of physical solutions and thus to preserve the hyperbolicity of the system.

For convenience of the illustration, we introduce the following transformation between nodal and modal representations of solutions.
Inside any element $\mathbf \Omega_i$, the solutions can be expressed as,
\begin{equation}
\begin{aligned}
    \mathbf u^\delta \simeq \mathbf u^N
    &= \sum_{j=1}^{N_p} \hat{\boldsymbol v}^\delta_{j} \mathbf \Phi \ell_{j}
    = \sum_{j=1}^{N_p} \sum_{k=0}^{N_c} \hat{\mathbf u}_{j,k}^{\delta} \ell_{j} \Phi_k \\
    &= \tilde{\mathbf u}^N = \sum_{j=0}^{N_p-1} \sum_{k=0}^{N_c} \tilde{\mathbf u}_{j,k}^{\delta} \Psi_{j} \Phi_k,
\end{aligned}
\label{eqn:modal solution}
\end{equation}
where the orthogonal polynomials $\{\Psi_j,\Phi_k\}$ are used in both reference physical and random space, with degrees $N_p-1$ and $N_c$, respectively.
The nodal and modal representations of gPC coefficients are related by the Vandermonde matrix,
\begin{equation}
    \mathcal V \hat{\mathbf u}^\delta = \tilde{\mathbf u}^\delta,
\end{equation}
where the entries of the Vandermonde matrix write,
\begin{equation}
    \mathcal V_{jk} = \Psi_k(\mathbf r_j).
    \label{eqn:vandermonde matrix}
\end{equation}

\subsection{Filter}

\subsubsection{Exponential filter}
\label{sec:exfilter}

The idea of filtering is to dampen the coefficients in the polynomial expansions.
Such damping effect is expected to vanish as the expansion term approaches infinity in the sense of consistency.
The exponential filter is arguably the most widely used filter for spectral methods \cite{hesthaven2007nodal,Hou2007}. It was recently used to reduce oscillations and increase convergence speed in kinetic equations \cite{Di2018,Fan2020a} as well as uncertainty quantification \cite{kusch2021oscillation,alldredge2021realizable}.
Given a one-dimensional modal solution $\tilde{\mathbf u}^N$, the exponential filtering takes the form
\begin{equation}
    \mathbf u^{*} = \mathcal F(\tilde{\mathbf u}^{N}) = \sum_{k=0}^N \lambda_k \tilde {\mathbf u}_{k} \Phi_k,
\end{equation}
where $\mathbf u^{*}$ is the post-filter solution.
The filter strength $\lambda$ is defined as,
\begin{equation}
    \lambda_k\left(\eta=\frac{k}{N}\right)=\left\{\begin{array}{ll}
    1, \quad 0 \leq \eta < \eta_{*}=\frac{N_{*}}{N}, \\
    \exp \left(-\alpha\Delta t\left(\left(\eta-\eta_{*}\right) /\left(1-\eta_{*}\right)\right)^{s}\right), \quad \eta_{*} \leq \eta \leq 1.
    \end{array}\right.
\end{equation}
Here, $N^* \geq 0$ represents a cutoff below which the modes are left untouched, e.g., $N^* = \frac{2}{3}N$ as recommended by \cite{Hou2007}. The exponent $s$ is an integer to be determined in specific examples, with $s=36$ in \cite{Hou2007,Koellermeier2020a}. The filter parameter $\alpha \geq 0$ is chosen as $\alpha = 36$ in \cite{Hou2007} to ensure that the last mode is fully damped up to machine precision. The choice of $\alpha$ largely depends on the application and several ways to choose appropriate filter parameters are discussed in detail in \cite{Fan2020a}. We refer to \ref{app} for more details and a parameter study of the filter as used in this work. The necessary parameter choices are an apparent drawback of the exponential filter but also allow for some flexibility in applications.

The filter operator $\mathcal F$ can be written as,
\begin{equation}
    \mathcal F(\tilde{\mathbf u}^{N}) = \lambda \circ \tilde{\mathbf u}^{N},
\end{equation}
where $\lambda = [\lambda_0,\lambda_1,\cdots,\lambda_N]^T$.

The above filter can be extended to multi-variate modal solutions in Eq.(\ref{eqn:modal solution}), i.e.,
%
\begin{equation}
\begin{aligned}
    &\mathbf u^{*} = \mathcal F(\mathbf u^{N}) = \sum_{j=0}^{N_p-1} \sum_{k=0}^{N_c} \lambda_{j,k} \tilde{\mathbf u}_{j,k}^{\delta} \Psi_{j} \Phi_k, \\
    &\lambda_{j,k}\left(\eta_1=\frac{j}{N_p},\eta_2=\frac{k}{N_c+1}\right)=\left\{\begin{array}{ll}
    1, \quad 0 \leq \eta < \eta_{*,\{1,2\}}, \\
    {\displaystyle \prod_{i=1}^2}\exp \left(-\alpha\Delta t\left(\left(\eta_i-\eta_{*,i}\right) /\left(1-\eta_{*,i}\right)\right)^{s}\right), \quad \text{else},
    \end{array}\right.
\end{aligned}
\end{equation}
where $\eta_{*,1}=\frac{N_{*}}{N_p}$ and $\eta_{*,2}=\frac{N_{*}}{N_c+1}$. Note that also the filter parameter and exponent can be made dependent on the dimension. 
The filter operator can be abbreviated again as,
\begin{equation}
    \mathcal F = \Lambda \circ,
\end{equation}
where the $\lambda_{j,k}$ are the entries of the matrix $\Lambda$.
It is noticeable that the filter operator can act on the nodal solution directly in practice, where the equivalent filter operator becomes,
\begin{equation}
    \mathcal F^* = \mathcal V \Lambda \circ \mathcal V^{-1},
\end{equation}
where $\mathcal V$ is the Vandermonde matrix defined in Eq.(\ref{eqn:vandermonde matrix}).

\subsubsection{L$^2$ filter}

As the spectral solution is dedicated to approximating the exact solution $\mathbf u^\delta$,
we can define the discrepancy between the approximation and the exact solutions based the norms of the solution matrix.
For example, the cost function of the $L^2$ norm can be written as,
\begin{equation}
    \mathcal{C}(\mathbf{u}^N):=\frac{1}{V} \int_{\mathbf \Omega} \int_{\mathbf \Upsilon}\left\|\mathbf u^\delta - \tilde{\mathbf u}^N \right\|^2_{L^2} \varrho(\mathbf r) \varrho(\mathbf z) d \mathbf r d \mathbf z,
    \label{eqn:gpc cost}
\end{equation}
where $V$ is the volume of the phase space and $\varrho$ denotes the probability density in the spatial and random domains.

The $L^2$ filter based on splines \cite{canuto1982approximation} regularizes the above error to mitigate oscillations.
A penalty term can be introduced into Eq.(\ref{eqn:gpc cost}),
\begin{equation}
\begin{aligned}
    \mathcal{C}_{\alpha}(\mathbf{u}^N)
    :=&\frac{1}{V} \int_{\mathbf \Omega} \int_{\mathbf \Upsilon}\left\|\mathbf u^\delta - \sum_{j=0}^{N_p-1}\sum_{k=0}^{N_c} \tilde{\mathbf u}_{j,k}^\delta \Psi_j \Phi_{k}\right\|_{L^2}^2 \varrho(\mathbf r) \varrho(\mathbf z) d \mathbf r d \mathbf z \\
    &+\int_{\mathbf \Omega} \int_{\mathbf \Upsilon} \left\| \alpha_1 \mathcal L_1 \sum_{j=0}^{N_p-1}\sum_{k=0}^{N_c} \tilde{\mathbf u}_{j,k}^\delta \Psi_j \Phi_{k} + \alpha_2 \mathcal L_2 \sum_{j=0}^{N_p-1}\sum_{k=0}^{N_c} \tilde{\mathbf u}_{j,k}^\delta \Psi_j \Phi_{k} \right\|^2_{L^2} \varrho(\mathbf r) \varrho(\mathbf z) d \mathbf r d \mathbf z,
    \label{eqn:gpc punish cost}
\end{aligned}
\end{equation}
where the operator $\mathcal L$ is used to punish the possible oscillations and $\alpha_{1,2} \in \mathcal R^+$ are the filter parameters.
A common choice of the penalty operator is 
\begin{equation}
    \mathcal L_i u(\bm y)=\partial_{y_i}\left((1-y_i^2)\partial_{y_i}u(\bm y)\right),
\end{equation}
where $\bm{y} = (y_1,\cdots,y_M)$ is an arbitrary vector-valued input. Note that the Legendre polynomials are eigenfunctions of this operator.
Differentiating Eq.(\ref{eqn:gpc punish cost}) with respect to the $L^2$ norm yields the optimal coefficients,
\begin{equation}
    \mathbf u^* = \sum_{j=0}^{N_p-1} \sum_{k=0}^{N_c} \tilde{\mathbf u}_{j,k}^{*} \Psi_{j} \Phi_k, \quad \mathbf{\tilde u}_{j,k}^* = \frac{\mathbf{\tilde u}_{j,k}^\delta}{1+\alpha_1 j^2(j+1)^2+\alpha_2 k^2(k+1)^2},
    \label{eqn:filter function}
\end{equation}
where $\mathbf{\hat u}_{j,k}^*$ denotes the coefficients after filtering.
As can be seen, the filter leaves the zeroth-order coefficients untouched and thus preserve the conservation of the expected value.

The filter parameters $\{\alpha_1,\alpha_2\}$ have yet to be determined.
If we specify the damping ratio of the last expansion term, the filter parameter can be obtained via,
\begin{equation}
    \alpha_1 = \frac{1}{\varepsilon_1 N_p^2 (N_p-1)^2}, \quad \alpha_2 = \frac{1}{\varepsilon_2 N_c^2 (N_c+1)^2},
    \label{eqn:l2 strength}
\end{equation}
where $\varepsilon_1$ and $\varepsilon_2$ denote the relative magnitudes of coefficients in the last expansion term with respect to spatial and random space.
Usually $\{\varepsilon_1,\varepsilon_2\}$ take higher values than for the exponential filter, where the last expansion term is dampened towards zero. Furthermore, note that if $N_p$ and $N_c$ tend to infinity, the above choice of the filter parameter ensures convergence, as the filtering effect vanishes in the limit.

\subsubsection{Lasso filter}

The cost function of the approximation solution can be defined on other norms, e.g., the $L^1$ norm.
In \cite{kusch2020filtered}, the filtering idea is combined with Lasso regression, and we can propose the following cost function in multi-dimensional space,
\begin{equation}
\begin{aligned}
    \mathcal{C}_{\alpha}(\mathbf u^*) := &\frac{1}{V} \int_{\mathbf \Omega} \int_{\mathbf \Upsilon}\left\|\mathbf u^\delta - \mathbf u^* \right\|^2_{L^2} \varrho(\mathbf r) \varrho(\mathbf z) d \mathbf r d \mathbf z \\
    &+ \frac{1}{V} \int_{\mathbf \Omega} \int_{\mathbf \Upsilon} \sum_{j=1}^{N_p-1}\sum_{k=1}^{N_c} \left(\alpha_1\left\Vert \mathcal L_1 \tilde{\mathbf u}_{j,k}^* \Psi_{j,k} \right\Vert_{1}+\alpha_2\left\Vert \mathcal L_2 \tilde{\mathbf u}_{j,k}^* \Psi_{j,k} \right\Vert_{1}\right) \varrho(\mathbf r) \varrho(\mathbf z) d \mathbf r d \mathbf z,
\label{eq:LassoFunctional}
\end{aligned}    
\end{equation}
where the penalty term is based on the $L^1$ norm and acts on the expansion term individually.

Conveniently, the above optimization problem has an analytic solution, therefore reducing computational costs significantly. The result follows from a straightforward extension of \cite[Theorem~1]{kusch2020filtered}:
\begin{theorem}
The minimizer of \eqref{eq:LassoFunctional} takes the form
\begin{align}\label{eq:LassoFilterFunction}
\mathbf{\tilde u}_{j,k}^* = \mathrm{ReLU} \left(1 - \frac{\alpha_1 j(j+1)\Vert \Psi_j\Phi_k \Vert_{1}}{\vert \tilde{\mathbf u}_{j,k}^\delta \vert}-\frac{\alpha_2 k(k+1)\Vert \Psi_j\Phi_k \Vert_{1}}{\vert \tilde{\mathbf u}_{j,k}^\delta \vert}\right) \mathbf{\tilde u}_{j,k}^\delta,
\end{align}
where ReLU is the rectified linear unit function and $\Vert\cdot\Vert_1$ denotes the L$^1$ norm.
\end{theorem}
\begin{proof}
For ease of presentation, we assume the solution to be scalar. Let us denote potential minimizers by $\bm{\alpha}\in\mathbb{R}^{(N_p-1)\times N_c}$. To minimize the cost functional \eqref{eq:LassoFunctional}, we need to determine the gradient. Since the cost function is not smooth, Lasso regression relies on the subdifferential \cite{tibshirani1996regression} instead of the gradient. The subdifferential with respect to the expansion coefficient $\alpha_{i,\ell}$ is denoted by $\partial_{i,\ell} \mathcal{C}_{\alpha}(\bm{v})$. When $v_{i,\ell} = 0$, we have
\begin{align}
    \partial_{i,\ell} \mathcal{C}_{\alpha}(\bm{v}) = \left\{ c_{i,\ell}(\bm{v},\gamma) : \gamma \in [-1,1] \right\}
\end{align}
where with $\tilde{\varrho}(r,z) := \varrho(r)\varrho(z)$ and $u^{\delta}_{i\ell}:=\int u^{\delta} \psi_i\phi_{\ell} \tilde{\varrho}\,drdz$ we have
\begin{align*}
    c_{i,\ell}(\bm{v},\gamma) := &\int \left( \sum_{j,k} v_{jk} \psi_j\phi_k - u^{\delta} \right) \psi_i\phi_{\ell} \tilde{\varrho}\,drdz+ \gamma \int \sum_{j,k} \left(\left\vert \alpha_1\mathcal{L}_1  v_{jk} \psi_j\phi_k \right\vert+\left\vert \alpha_2\mathcal{L}_2  v_{jk} \psi_j\phi_k \right\vert\right) \tilde{\varrho}\,drdz\\
    =&v_{i\ell} - u^{\delta}_{i\ell} + \gamma \int \left(\left\vert \alpha_1 i(i+1)  \psi_i\phi_{\ell} \right\vert+\left\vert \alpha_2\ell(\ell+1)  \psi_i\phi_{\ell} \right\vert\right) \tilde{\varrho}\,drdz.
\end{align*}
To have optimality, we need $0\in \partial_{i,\ell} \mathcal{C}_{\alpha}(\bm{v})$, i.e., if 
\begin{align}\label{eq:intervalLasso}
    u^{\delta}_{i\ell}\in\left[-\left(\alpha_1 i(i+1) + \alpha_2 \ell(\ell+1)\right)  \Vert\psi_i\phi_{\ell} \Vert_1,\left(\alpha_1 i(i+1) + \alpha_2 \ell(\ell+1)\right)  \Vert\psi_i\phi_{\ell} \Vert_1\right]
\end{align}
we must set $v_{i,\ell}$ to zero. If $u^{\delta}_{i\ell}$ does not fulfill \eqref{eq:intervalLasso}, we know that $v_{i,\ell}\neq 0$ and the cost function is differentiable. Then, the gradient can be computed and the optimality condition is simply given by
\begin{align*}
    \partial_{i,\ell} \mathcal{C}_{\alpha}(\bm{v}) = v_{i\ell} - u^{\delta}_{i\ell} +  \text{sign}(v_{i,\ell})\left(\alpha_1 i(i+1) + \alpha_2 \ell(\ell+1)\right)\Vert\psi_i\phi_{\ell} \Vert_1 \stackrel{!}{=} 0.
\end{align*}
Hence, if \eqref{eq:intervalLasso} does not hold, we have
\begin{align*}
    v_{i\ell} = u^{\delta}_{i\ell} - \text{sign}(v_{i,\ell})\left(\alpha_1 i(i+1) + \alpha_2 \ell(\ell+1)\right)\Vert\psi_i\phi_{\ell} \Vert_1.
\end{align*}
Following the proof of \cite[Theorem~1]{kusch2020filtered}, this can be written down compactly as \eqref{eq:LassoFilterFunction}.
\end{proof}

The Lasso filter yields an automated and adaptive strategy to pick an adequate filter parameter. Following \cite{kusch2020filtered}, we wish to choose the filter parameter, such that no information is lost through the imposed polynomial truncation. A likely scenario which achieves this goal is when the filter sets the highest expansion coefficients to zero. To ensure that the filtered coefficients $\tilde{\mathbf u}_{N_p-1,0}^*$ and $\tilde{\mathbf u}_{0,N_c}^*$ are zero, this leads to,
\begin{equation}
\begin{aligned}
    &\alpha_1 = \frac{\| \tilde{\mathbf u}_{N_p-1,0}^\delta \|_{L^1}}{ N_p(N_p-1) \Vert \Psi_{N_p-1}\Phi_{0} \Vert_{1}},\\
    &\alpha_2 = \frac{\| \tilde{\mathbf u}_{0,N_c}^\delta \|_{L^1}}{ N_c(N_c+1) \Vert \Psi_{0}\Phi_{N_c} \Vert_{1}}.
\end{aligned}
\label{eq:fStrength}
\end{equation}

\subsection{Discontinuity detector}\label{sec:detector}

In contrast to the Lasso filter, the L$^2$ filter is used globally each step or every few steps during the simulation.
This may lead to a loss of accuracy in smooth regions, where the solution structure has been well captured by the polynomial expansions.
A better strategy would be that appropriate numerical dissipation is injected only when it is needed.
This requires a proper detection of discontinuous solutions.
Here we follow the sensor for discontinuities proposed in \cite{persson2006sub} for the discontinuous Galerkin methods. The sensor has been used in \cite{rajput2020master} for the filtered stochastic-Galerkin method.

Let us consider the modal solution in the element $\mathbf \Omega_i$,
\begin{equation}
    \mathbf u^N
    = \sum_{j=0}^{N_p-1} \sum_{k=0}^{N_c} \tilde{\mathbf u}_{j,k}^{\delta} \Psi_{j} \Phi_k,
\end{equation}
where $N_p$ is the number of solution points, and $N_c$ is the degree of polynomial chaos in the random space.
In the smooth region, the coefficients $\tilde{\mathbf u}_{j,k}^{\delta}$ are expected to decrease quickly with increasing polynomial order.
Therefore, a slope indicator can be defined as
\begin{equation}
    \mathbf S_e = \frac{\langle \mathbf u^N- \mathbf u^{N-1}, \mathbf u^N- \mathbf u^{N-1} \rangle}{\langle \mathbf u^N, \mathbf u^N \rangle},
\end{equation}
where $\mathbf u^{N-1}$ denotes a truncated expansion of the same solution at order $N-1$.
The indicator $\mathbf S_e$ can be a non-negative number for scalar transport equations, or a vector for a system of equations.
We extract the first state of $\mathbf S_e$ and define it as $S_e$.
A discontinuity detector can be formulated as,
\begin{equation}
\begin{aligned}
    & \theta =\left\{\begin{array}{ll}
    1,  \quad & s_{e}<s_{0}-\kappa, \\
    \frac{1}{2}\left(1-\sin \frac{\pi\left(s_{e}-s_{0}\right)}{2 \kappa}\right), \quad & s_{0}-\kappa \leq s_{e} \leq s_{0}+\kappa, \\
    0, & s_{e}>s_{0}+\kappa.
    \end{array}\right. \\
    \\
    & \theta < 0.99 \ \longrightarrow \ \text{discontinuity},
\end{aligned}
\label{eqn:detector}
\end{equation}
where $s_e = \log_{10}(S_e)$.
The parameter $s_0$ is chosen to be inversely proportional to the polynomial degree, and $\kappa$ needs to be sufficiently large to obtain a sharp and non-oscillating solution profile.

\subsection{Positivity preserving limiter}

The use of filters suffices to mitigate the Gibbs phenomenon and thus stabilizes the numerical computation.
However, it does not necessarily preserve the realizability of physical solutions, e.g., the non-negative density and temperature in the Euler equations.
It is feasible to apply filters either with sufficiently strong filter parameter \cite{alldredge2021realizable} or successively \cite{xiao2021stochastic}, while the excess introduction of artificial dissipation may cause a severe loss of accuracy or even break the physical structure.
In this paper, we adopt a slope limiter in conjunction with the filter to preserve the positivity of realizable solutions.
The idea of limiting the solution slopes comes naturally from the development of high-order methods, e.g., the discontinuous Galerkin method \cite{burbeau2001problem} and the flux reconstruction method \cite{vandenhoeck2019implicit}. 
We extend the limiter proposed in \cite{vandenhoeck2019implicit} to multi-dimensional spatial-random space.
A similar strategy has been applied in \cite{durrwachter2020hyperbolicity}, which extends the limiter in random space \cite{schlachter2018hyperbolicity} under the DG framework.

For clarity, we take the Euler equations as an example.
In the solution algorithm, we first evaluate the polynomial chaos at quadrature points in the random space and get a fully nodal representation.
The mean density $\bar \rho_i$ and mean pressure $\bar p_i$ are calculated in each element $\mathbf \Omega_i$.
For an interpolation higher than $P_1$, the local extrema of density and pressure can emerge at any point in the element, and thus we need to detect the minimum value among both solution points and flux points.
This step can be done together with the Lagrange interpolation for the interface flux calculation in Eq.(\ref{eqn:interactive flux}).
As we demand positivity of both density and pressure, the limiter is turned on when the following condition is satisfied,
\begin{equation}
    \min(\rho_\mathrm{min},p_\mathrm{min}) < \epsilon,
\end{equation}
where the small parameter $\epsilon$ is defined via,
\begin{equation}
    \epsilon = \min(10^{-8}, \bar \rho_i, \bar p_i).
\end{equation}

The density value at the $j$-th solution point in the physical space and the $k$-th quadrature point in the random space can be reconstructed with limited slopes as,
\begin{equation}
    {\rho}^\star_{i,j,k}=\beta_{1}\left(\rho_{i,j,k}-\bar{\rho}_i\right)+\bar{\rho}_i, 
    \quad \beta_{1}=\min \left(\frac{\bar{\rho}_i-\epsilon}{\bar{\rho}_i-\rho_{\min }}, 1\right).
\end{equation}
In this way, the density values and slopes in the element are limited.

We then construct an intermediate state ${\mathbf u}^\star=(\rho^\star, \rho \mathbf v, \rho \mathbf E)$.
If the positivity of pressure is not satisfied, i.e., $p^\star<\epsilon$,
the following nonlinear equation is solved at all the solution and flux points,
\begin{equation}
    p\left(\beta_{l}\left({\mathbf{u}}_{i,l,k}^{\star}-\overline{\mathbf{u}}_i\right)+\overline{\mathbf{u}}_i\right)=\epsilon,
\end{equation}
where the corresponding slope restriction $\beta_l$ can be obtained at different locations.
The final limited solution at the $j$-th solution point and the $k$-th quadrature point is computed by
\begin{equation}
    {\mathbf{u}}_{i,j,k}^+=\beta_{2}\left({\mathbf{u}}_{i,j,k}^\star-\bar{\mathbf{u}}_i\right)+\bar{\mathbf{u}}_i, \quad \beta_{2}=\min _{l}\left(\beta_{l}\right).
\end{equation}
This scheme guarantees that the density and pressure stay positive at the solution and flux points.
Let us now write down the fully discretized scheme. For sake of readability, we assume a forward Euler time discretization. However, other discretizations are possible.
Considering the solution $\mathbf u_i^n$ and its average $\bar{\mathbf u}_i^n$ at time step $t^n$ inside the standard element $\mathbf \Omega_i$, the solution algorithm yields,
\begin{equation}\label{eq:schemefullDiscr}
\begin{aligned}
    & {\mathbf{u}}_{i}^{n+1/3}=\beta_{2}\left({\mathbf{u}}_{i}^{n}-\bar{\mathbf{u}}_i^n\right)+\bar{\mathbf{u}}_i^n, \\
    & \mathbf{u}_{i}^{n+2/3} =\mathbf{u}_{i}^{n+1/3} -\Delta t \nabla_{\mathbf r} \cdot {\mathbf f}^{n+1/3} (\mathbf r_j),\\
    & \mathbf{u}_{i}^{n+1} = {\mathcal{F}}(\mathbf{u}_{i}^{n+2/3}),\\
\end{aligned}
\end{equation}
Here, the filtering step is denoted by ${\mathcal{F}}$. For $\Delta x,\Delta t \rightarrow 0$ and without filtering, the scheme \eqref{eq:schemefullDiscr} solves the following equations,
\begin{equation}\label{eq:schemeConsistent}
\begin{aligned}
    &\partial_{t} \hat{\mathbf{u}}_{k} + \langle \nabla \cdot \mathbf{f}\left(\mathbf{\widetilde u}^{N}\right) \Phi_{k} \rangle =\mathbf{0},\\
    &\mathbf{\widetilde u}^{N}(t,\mathbf x,\mathbf z)=\beta_{2}\left(\mathbf{u}^{N}(t,\mathbf x,\mathbf z)-\bar{\mathbf{u}}_i(t,\mathbf x)\right)+\bar{\mathbf{u}}_i(t,\mathbf x).
\end{aligned}
\end{equation}
Following \cite[Theorem~2.1]{wu2017stochastic}, the above stochastic Galerkin system \eqref{eq:schemeConsistent} is hyperbolic. I.e., the method presented in this work provides a bound-preserving high-order discretization of the hyperbolic moment system.
For $\Delta x,\Delta t \rightarrow 0$, the solution algorithm is consistent with the hyperbolicity-preserving SG method \cite{schlachter2018hyperbolicity}.

\section{Numerical Experiments}\label{sec:numerical experiments}

In this section, we will conduct numerical experiments to validate the current scheme.
The dimensionless variables are introduced as follows, 
\begin{equation*}
    \tilde{\mathbf{x}}=\frac{\mathbf{x}}{L_{0}}, \ \tilde{t}=\frac{t}{L_{0} / V_{0}}, \
    \tilde{\mathbf{u}}=\frac{\mathbf{u}}{U_{0}}, 
\end{equation*}
where $L_0$ is the reference length, $V_0$ is the reference speed and $U_0$ denotes the reference conservative variables.
For brevity, we drop the tilde notation to denote dimensionless variables henceforth.

\subsection{Advection equation}
First we study the convergence order of the current scheme. 
The one-dimensional wave propagation problem with random initial input is used as the validation case, i.e.,
\begin{equation*}
    \partial_t u + a\partial_x u = 0, \quad u(t=0,x,z)=\xi(z) \sin(\pi x).
\end{equation*}
The exact solution follows,
\begin{equation*}
    u(t,x,z)=\xi(z) \sin(\pi (x-at)).
\end{equation*}
The detailed computational setup is recorded in Table \ref{tab:wave}, where $\mathcal U$ denotes the uniform distribution.
\begin{table}[htbp]
	\caption{Computational setup of wave propagation problem.} 
	\centering
	\begin{tabular}{lllllll} 
		\hline
		$t$ & $x$ & $z$ & $N_x$ & Points & $N_p$ & Correction \\
		$(0,50]$ & $[-1,1]$ & $[-1,1]$ & $[5,40]$ & Legendre & $[3,4]$ & Radau \\ 
		\hline
		$\xi$ & gPC & $N_c$ & $N_q$ & $a$ & Flux & Integrator \\ 
		$\mathcal U(0.9,1.1)$ & Legendre & 5 & 9 & 1 & Lax–Friedrichs & RK4 \\ 
		\hline
		Boundary & CFL \\
		Periodic & 0.1 \\
		\hline
	\end{tabular} 
	\label{tab:wave}
\end{table}

\begin{figure}
    \centering
    \begin{minipage}{0.49\textwidth}
        \includegraphics[width=\textwidth]{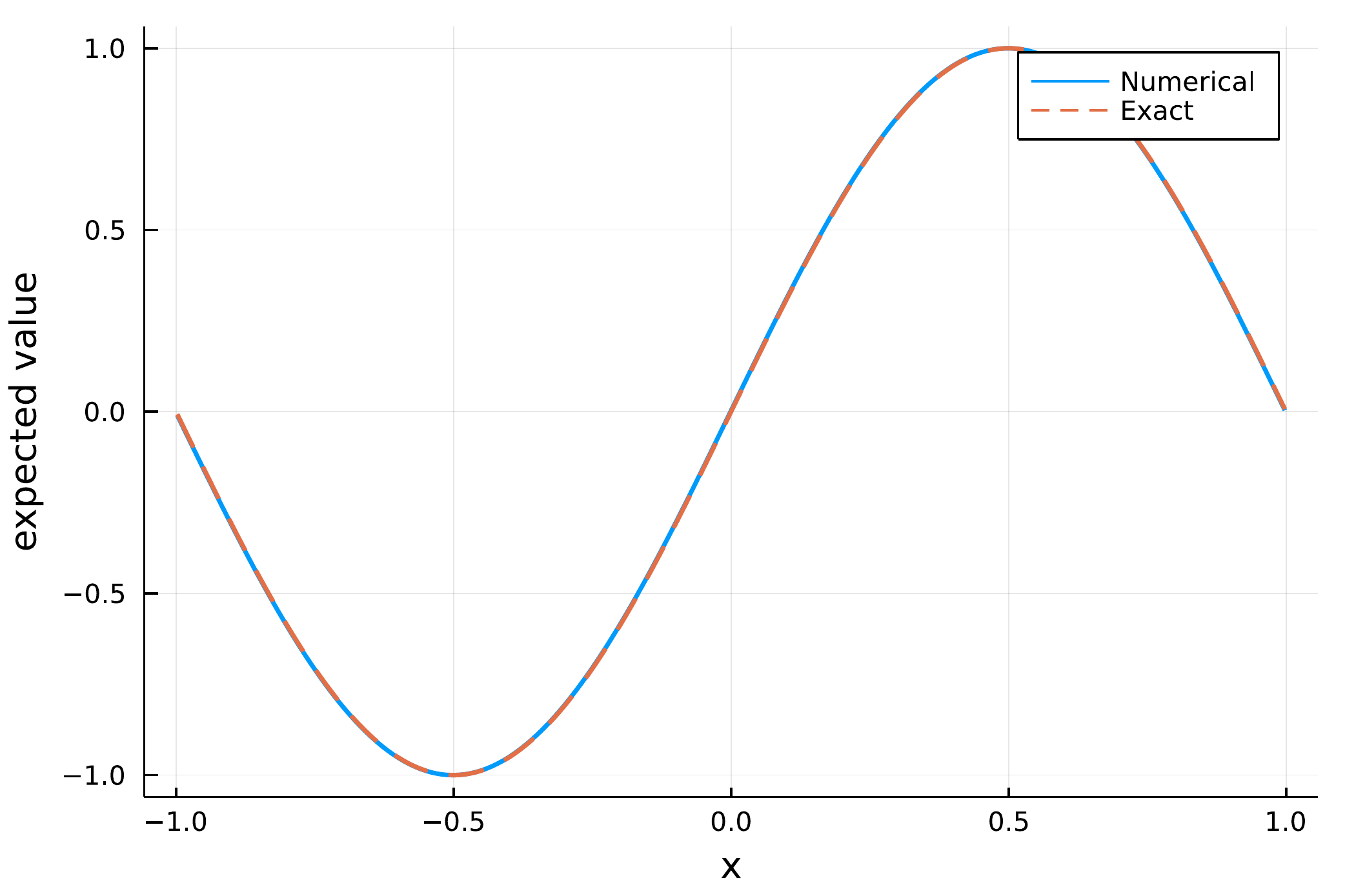}
    \end{minipage}
    \begin{minipage}{0.49\textwidth}
        \includegraphics[width=\textwidth]{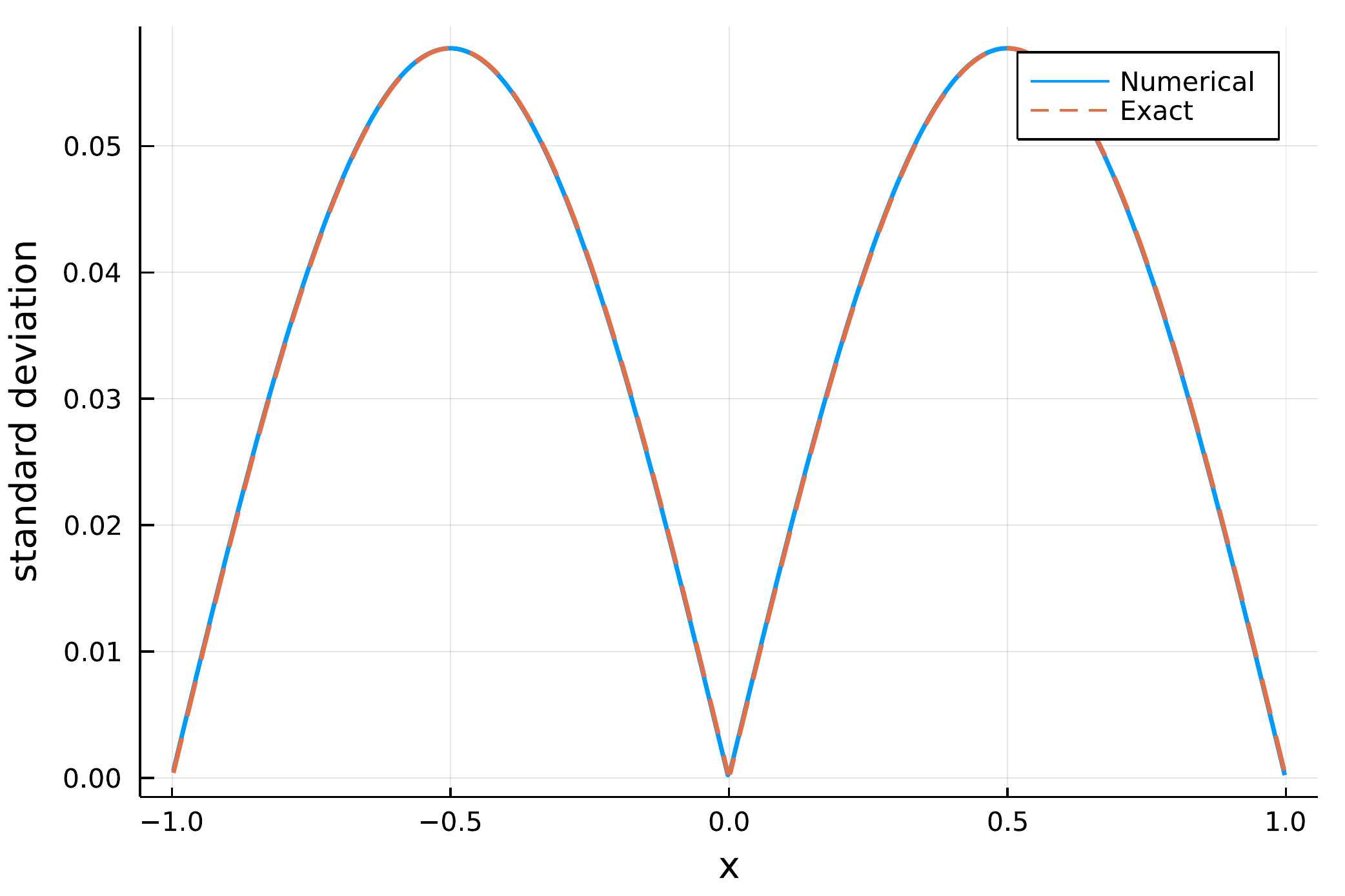}
    \end{minipage}
    \caption{The expected value and standard deviation of wave propagation problem with $N_x=40$ at $t=50$.}
    \label{fig:advection}
\end{figure}

The Lagrange polynomials of degree 2 and 3 are constructed in the computation, resulting in third- and fourth-order schemes, respectively.
Different number of elements from $N_x=5$ to $N_x=40$ are used to compute the numerical solutions.
Following the criterion in Eq.(\ref{eqn:detector}), the filter is turned off automatically in this case.
Table \ref{tab:wave accuracy 1} and \ref{tab:wave accuracy 2} list the numerical errors and orders of convergence.
It is clear that the current method preserves the desired accuracy.
Fig. \ref{fig:advection} shows the expected value and standard deviation of the transport scalar $u$ at $t=50$ with 40 elements and 3 collocation points inside each cell.
As shown, the long time behavior of the stochastic advection system is well captured.

\begin{table}[htbp]
	\caption{Errors and convergences in the wave propagation problem.} 
	\centering
	\begin{tabular}{lllllll} 
		\hline
		$\Delta x$ & $L^1$ error & Order & $L^2$ error & Order & $L^\infty$ error & Order \\ 
		\hline
		0.5 & 5.941757E-2 & & 1.960147E-2 & & 7.945011E-3 &  \\
        0.25 & 6.901634E-3 & 3.11 & 1.563457E-3 & 3.65 & 4.915774E-4 & 4.01 \\
        0.125 & 8.419116E-4 & 3.04 & 1.354328E-4 & 3.53 & 3.121555E-5 & 3.98 \\
        0.0625 & 1.045722E-4 & 3.01 & 1.191615E-5 & 3.51 & 1.965592E-6 & 3.99 \\
        0.03125 & 1.304139E-5 & 3.00 & 1.050922E-6 & 3.50 & 1.231935E-7 & 4.00 \\
		\hline
	\end{tabular} 
	\label{tab:wave accuracy 1}
\end{table}

\begin{table}[htbp]
	\caption{Errors and convergences in the wave propagation problem.} 
	\centering
	\begin{tabular}{lllllll} 
		\hline
		$\Delta x$ & $L^1$ error & Order & $L^2$ error & Order & $L^\infty$ error & Order \\ 
		\hline
		0.5 & 7.184865E-3 & & 1.912045E-3 & & 6.542437E-4 &  \\
        0.25 & 4.177470E-4 & 4.10 & 8.187982E-5 & 4.55 & 2.197287E-5 & 4.90 \\
        0.125 & 2.583430E-5 & 4.02 & 3.624978E-6 & 4.50 & 7.261641E-7 & 4.92 \\
        0.0625 & 1.619619E-6 & 4.00 & 1.635994E-7 & 4.47 & 2.314494E-8 & 4.97 \\
        0.03125 & 1.015061E-7 & 4.00 & 7.516612E-9 & 4.44 & 7.480234E-10 & 4.95 \\
		\hline
	\end{tabular} 
	\label{tab:wave accuracy 2}
\end{table}

\subsection{Inviscid Burgers' equation}

Now let us shift our attention from capturing smooth solutions to tackling the problems where resolved and unresolved regions coexist.
Following \cite{Poette2009}, we consider the inviscid Burgers' equation under stochastic initial condition,
\begin{equation}
\begin{aligned}
    & \partial_t u + u\partial_x u = 0, \\
    & u(t=0,x,z):= \begin{cases}u_{L}, & \text { if } x<x_{0}+ \xi z, \\ u_{L}+\frac{u_{R}-u_{L}}{x_{0}-x_{1}}\left(x_{0}+ \xi z-x\right), & \text { if } x \in\left[x_{0}+ \xi z, x_{1}+ \xi z\right], \\ u_{R}, & \text { else }.\end{cases}
\end{aligned}
\end{equation}
This test case presents a forming shock. The initially continuous solution profile moves through the physical domain and thereby forms an discontinuity. The detailed computational setup can be found in Table \ref{tab:burgers}, where the integrator denotes the Runge–Kutta pairs of order 5 (4) proposed by Tsitouras \cite{tsitouras2011runge} and $\{\varepsilon_1,\varepsilon_2\}$ are the parameters used to define the filter parameters in Eq.(\ref{eqn:l2 strength}). Note that the Lasso filter does not require these filter parameters as all parameters are picked automatically. For the $L^2$ filter, a parameter study has been conducted to determine adequate values.
%
\begin{table}[htbp]
	\caption{Computational setup of Burgers shock problem.} 
	\centering
	\begin{tabular}{lllllll} 
		\hline
		$t$ & $x$ & $z$ & $N_x$ & Points & $N_p$ \\
		$(0,0.1]$ & $[0,3]$ & $[-1,1]$ & $100$ & Legendre & $[4,6]$ \\ 
		\hline 
		Correction & $u_L$ & $u_R$ & $x_0$ & $x_1$ & $\xi$  \\ 
		Radau & 11 & 1 & 0.5 & 1.5 & $0.2$ \\
		\hline
		gPC & $N_c$ & $N_q$ & Flux & Integrator & Boundary \\
		Legendre & 9 & 17 & Lax–Friedrichs & Tsitouras 5(4) & Dirichlet \\
		\hline
		CFL & $s_0$ & $\kappa$ & $\varepsilon_1$ & $\varepsilon_2$ & $\alpha$ & $s$ \\
		0.1 & $-2 \log (N_p-1)$ & 4 & 0.6 & 0.6 & 36 & 3 \\
		\hline
	\end{tabular} 
	\label{tab:burgers}
\end{table}

\begin{figure}
    \centering
    \begin{minipage}{0.49\textwidth}
        \includegraphics[width=\textwidth]{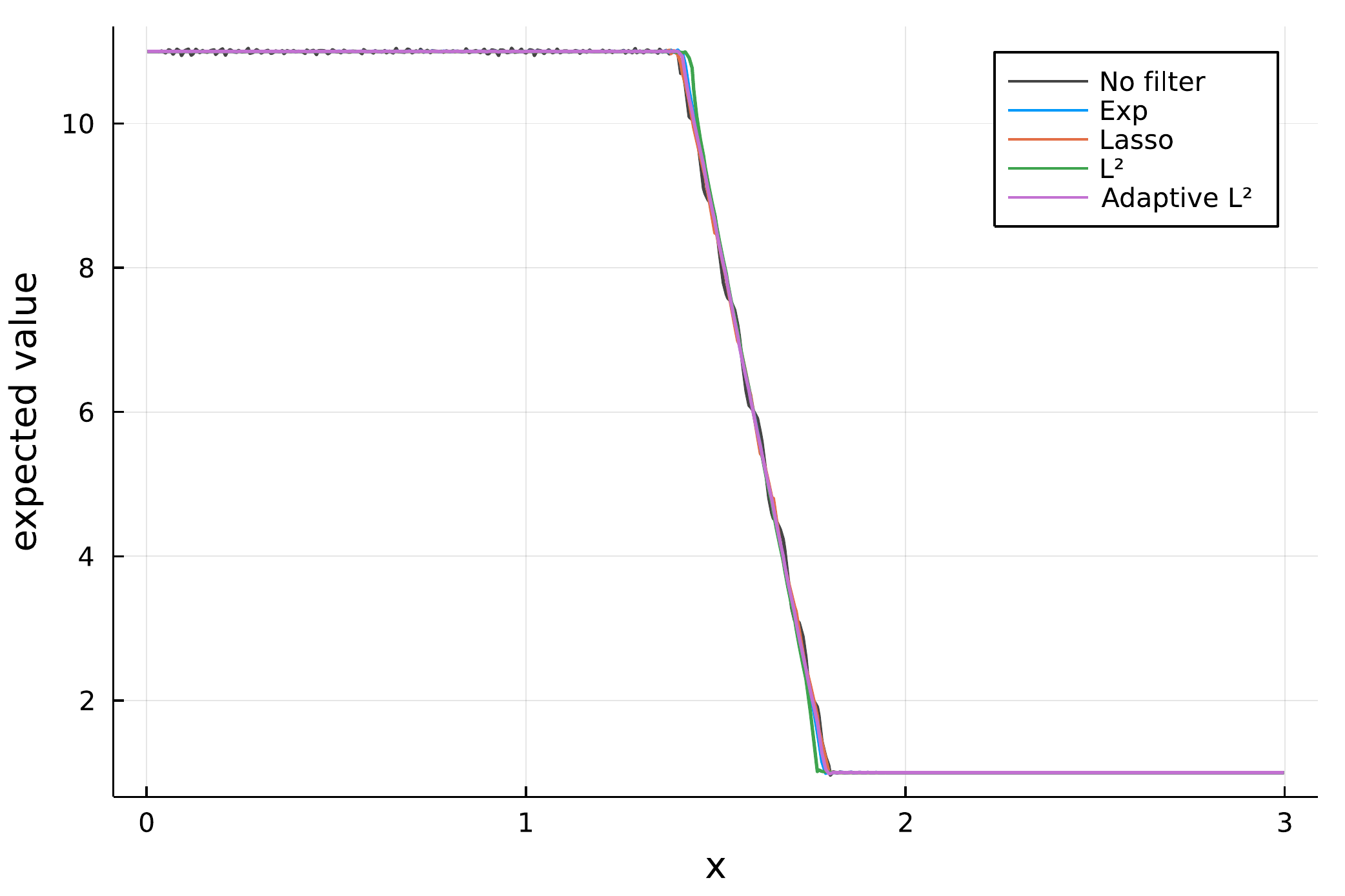}
    \end{minipage}
    \begin{minipage}{0.49\textwidth}
        \includegraphics[width=\textwidth]{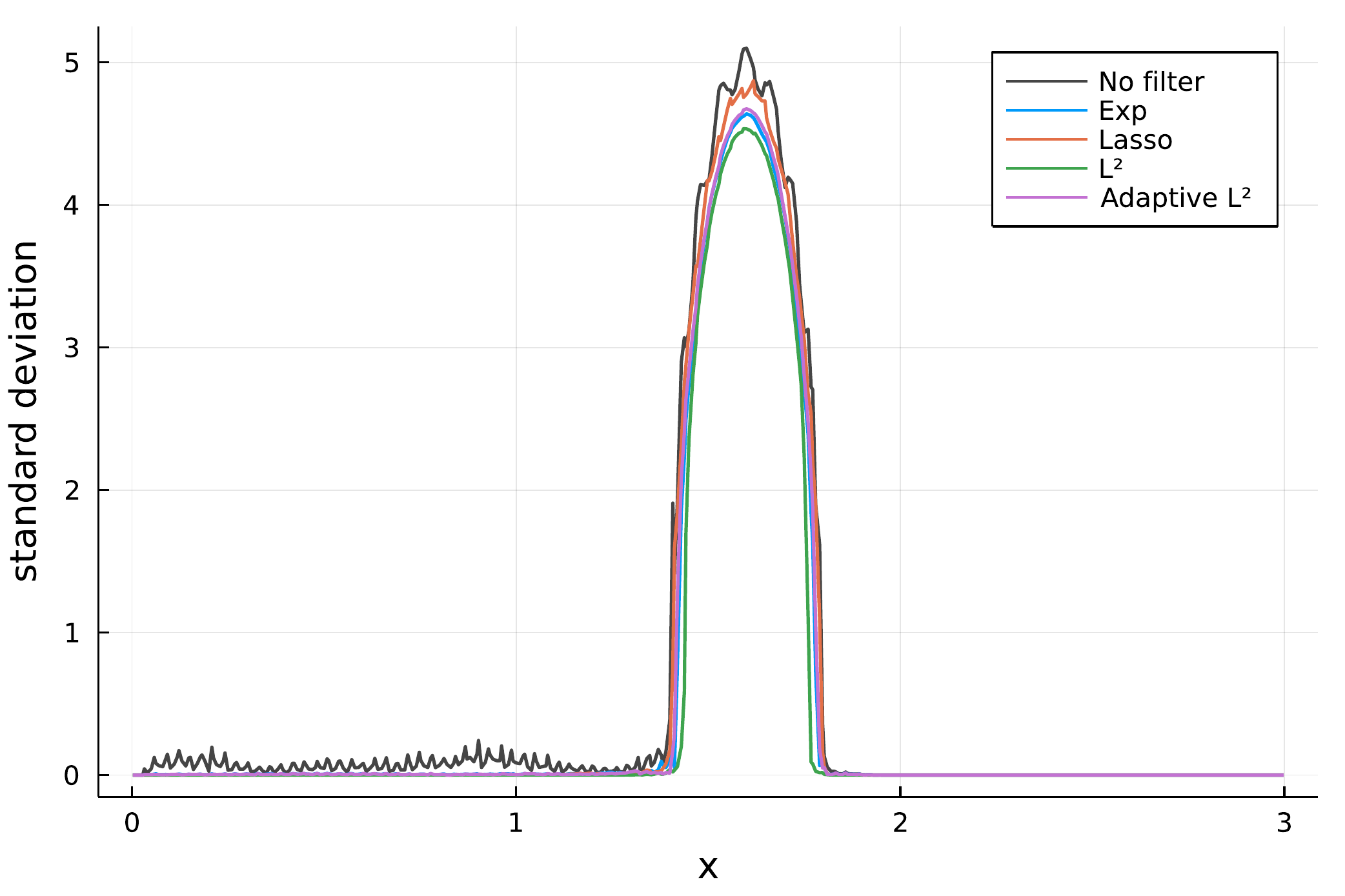}
    \end{minipage}
    \caption{The expected value and standard deviation of fourth-order Burgers' solutions at $t=0.1$ with different filters.}
    \label{fig:burgers4}
\end{figure}

\begin{figure}
    \centering
    \begin{minipage}{0.49\textwidth}
        \includegraphics[width=\textwidth]{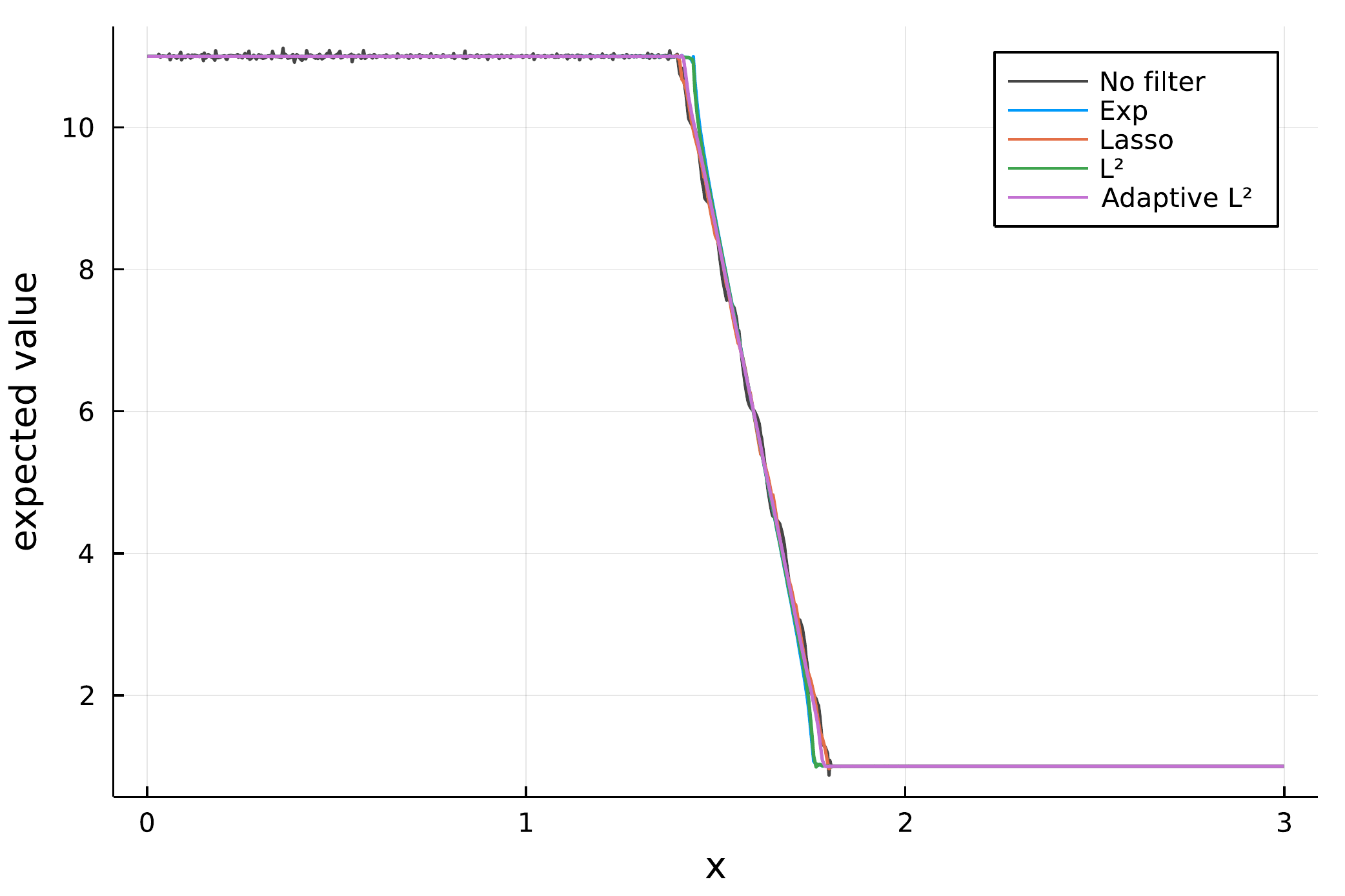}
    \end{minipage}
    \begin{minipage}{0.49\textwidth}
        \includegraphics[width=\textwidth]{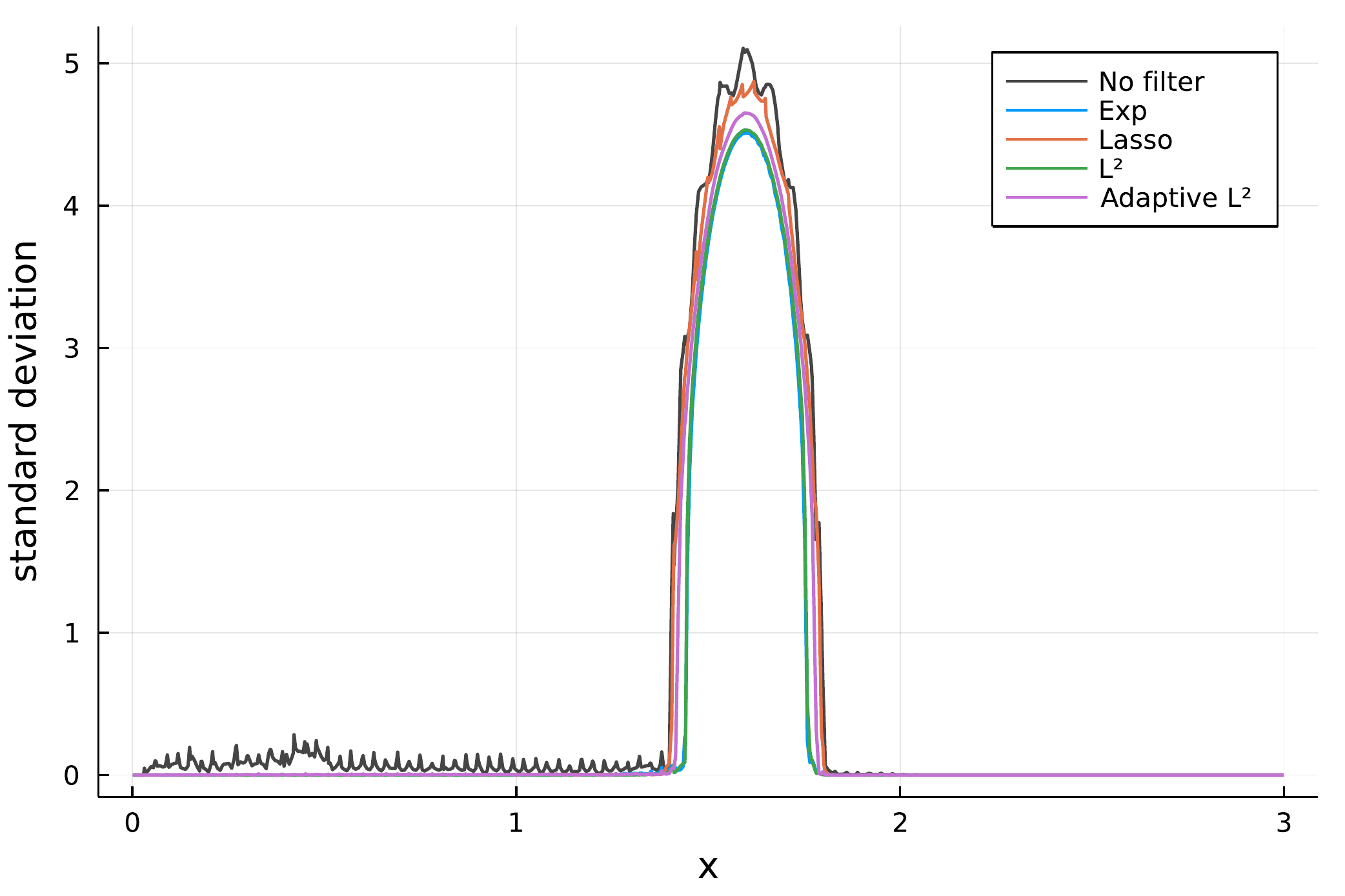}
    \end{minipage}
    \caption{The expected value and standard deviation of sixth-order Burgers' solutions at $t=0.1$ with different filters.} 
    \label{fig:burgers6}
\end{figure}

Fig. \ref{fig:burgers4} and \ref{fig:burgers6} show the profiles of expected value and standard deviation at $t=0.1$ from fourth and sixth order schemes, respectively, with 100 elements.
We compare the performance of different filters in this test case.
For the standard SG method, the Gibbs phenomenon results in spurious oscillations.
Compared to the expectation value, the variance is more sensitive and presents much stronger artifacts.
As is shown, all the filters help mitigate that in the upstream region.
In the shock region, all filters reduce oscillations, which the Lasso filter introducing the least numerical dissipation. The exponential filter and the $L^2$ filter show visibly more diffusive behavior.
This introduction of numerical dissipation inevitably reduces the peak value of the standard deviations.
However, benefiting from the discontinuity detector in section \ref{sec:detector}, the adaptive $L^2$ filter results in a significantly sharper profile while maintaining the robustness of the solution.
This numerical experiment demonstrates the leading performance of the Lasso filter and the adaptive filter, and thus we continue with them from now on.


\subsection{Sod shock tube}

We then turn to the Riemann problem in one-dimensional Euler equations,
\begin{equation}
    \frac{\partial}{\partial t}\left(\begin{array}{c}
    \rho \\
    \rho U \\
    \rho E
    \end{array}\right)+\frac{\partial}{\partial x} \left(\begin{array}{c}
    \rho {U} \\
    \rho {U}^2 +p \\
    (\rho E + p)U
    \end{array}\right)=\left(\begin{array}{c}
    0 \\
    0 \\
    0
    \end{array}\right).
\end{equation}
For the Sod shock tube, the uncertainties are introduced by the stochastic initial conditions, i.e.,  
\begin{equation}
\begin{aligned}
    & \mathbf v(t=0,x,z):= \left(\begin{array}{c}
    \rho \\
    U \\
    p 
    \end{array}\right) = \begin{cases} \mathbf v_{L}, & x<x_c, \\ \mathbf v_{R}, & x \geq x_c ,\end{cases}.
\end{aligned}
\end{equation}
Following \cite{xiao2021stochastic}, we consider two types of initial discontinuities. 
The first case employs stochastic density in the left-hand side,
\begin{equation}
\begin{aligned}
    \mathbf v_L = 
    \left(\begin{array}{c}
    \xi \\
    0 \\
    1
    \end{array}\right), \quad
    \mathbf v_R = 
    \left(\begin{array}{c}
    0.125 \\
    0 \\
    0.1
    \end{array}\right), \quad x_c = 0.5,
\end{aligned}
\end{equation}
while the location of initial discontinuity is stochastic in the second case, i.e.,
\begin{equation}
\begin{aligned}
    \mathbf v_L = 
    \left(\begin{array}{c}
    1 \\
    0 \\
    1
    \end{array}\right), \quad
    \mathbf v_R = 
    \left(\begin{array}{c}
    0.125 \\
    0 \\
    0.1
    \end{array}\right), \quad x_c = 0.5 + \sigma z.
\end{aligned}
\end{equation}
The second case is more challenging since the discontinuity is introduced in both physical and random space.
As discussed in \cite{poette2009uncertainty}, a negative density or temperature induced by the gPC
expansions may even lead to the failure of the solver at the first iterative step.
The detailed computational setup can be found in Table \ref{tab:euler}.
\begin{table}[htbp]
	\caption{Computational setup of Sod shock tube problem.} 
	\centering
	\begin{tabular}{lllllll} 
		\hline
		$t$ & $x$ & $z$ & $N_x$ & Points & $N_p$ & Correction \\
		$(0,0.15]$ & $[0,1]$ & $[-1,1]$ & $100$ & Legendre & $3$ & Radau \\ 
		\hline 
		$\xi$ & $\sigma$ & gPC & $N_c$ & $N_q$ & Flux & Integrator  \\ 
		$\mathcal U(0.9,1.1)$ & $0.05$ & Legendre & 9 & 17 & HLL & Bogacki-Shampine \\ 
		\hline
		Boundary & CFL & Filter & $s_0$ & $\kappa$ & $\varepsilon_1$ & $\varepsilon_2$ \\
		Dirichlet & 0.1 & (Lasso, $L^2$) & $-3 \log (N_p-1)$ & 4 & 0.6 & $(1, 0.6)$ \\
		\hline
	\end{tabular} 
	\label{tab:euler}
\end{table}

\begin{figure}
    \centering
    \begin{minipage}{0.49\textwidth}
        \includegraphics[width=\textwidth]{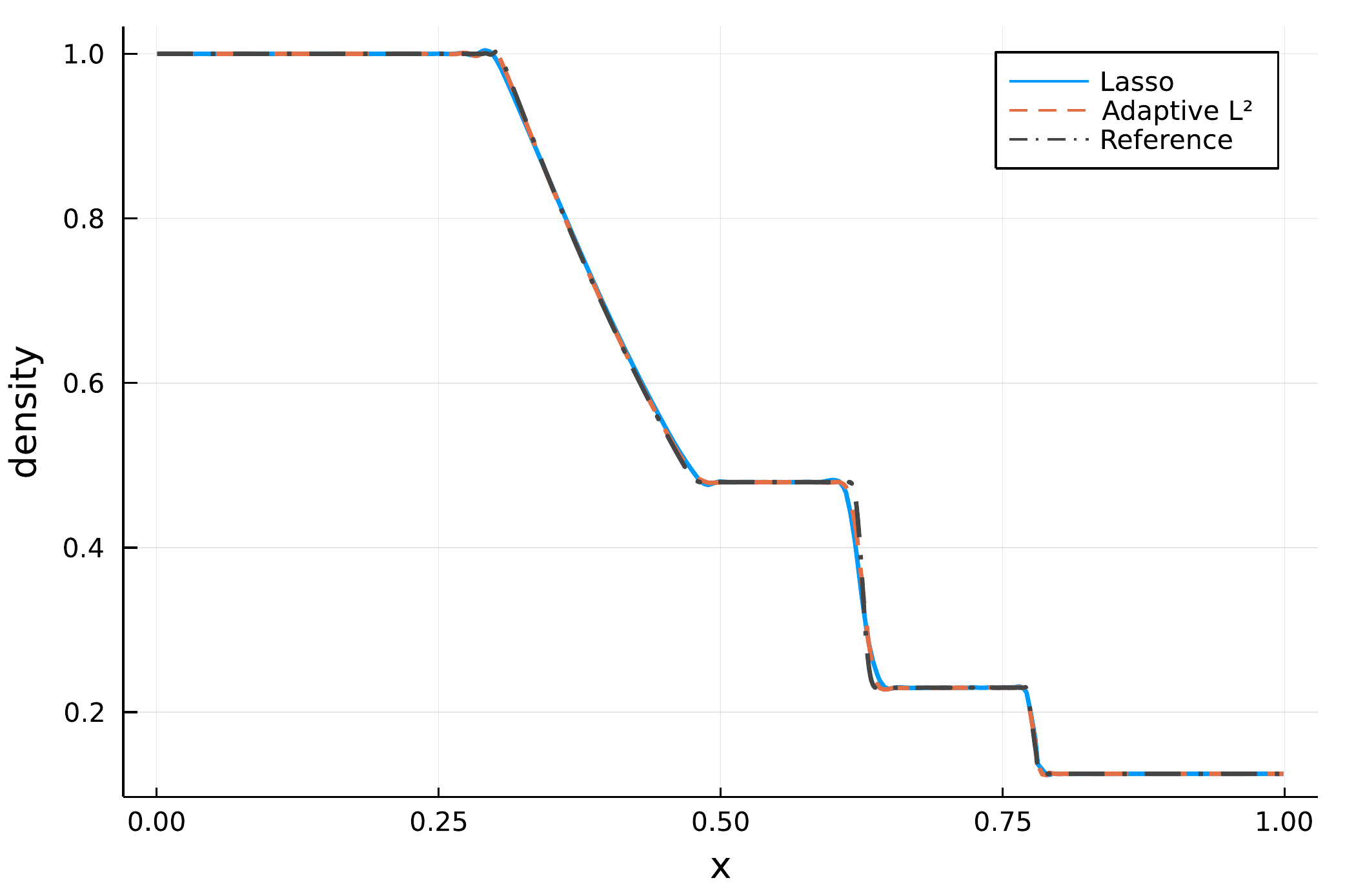}
    \end{minipage}
    \begin{minipage}{0.49\textwidth}
        \includegraphics[width=\textwidth]{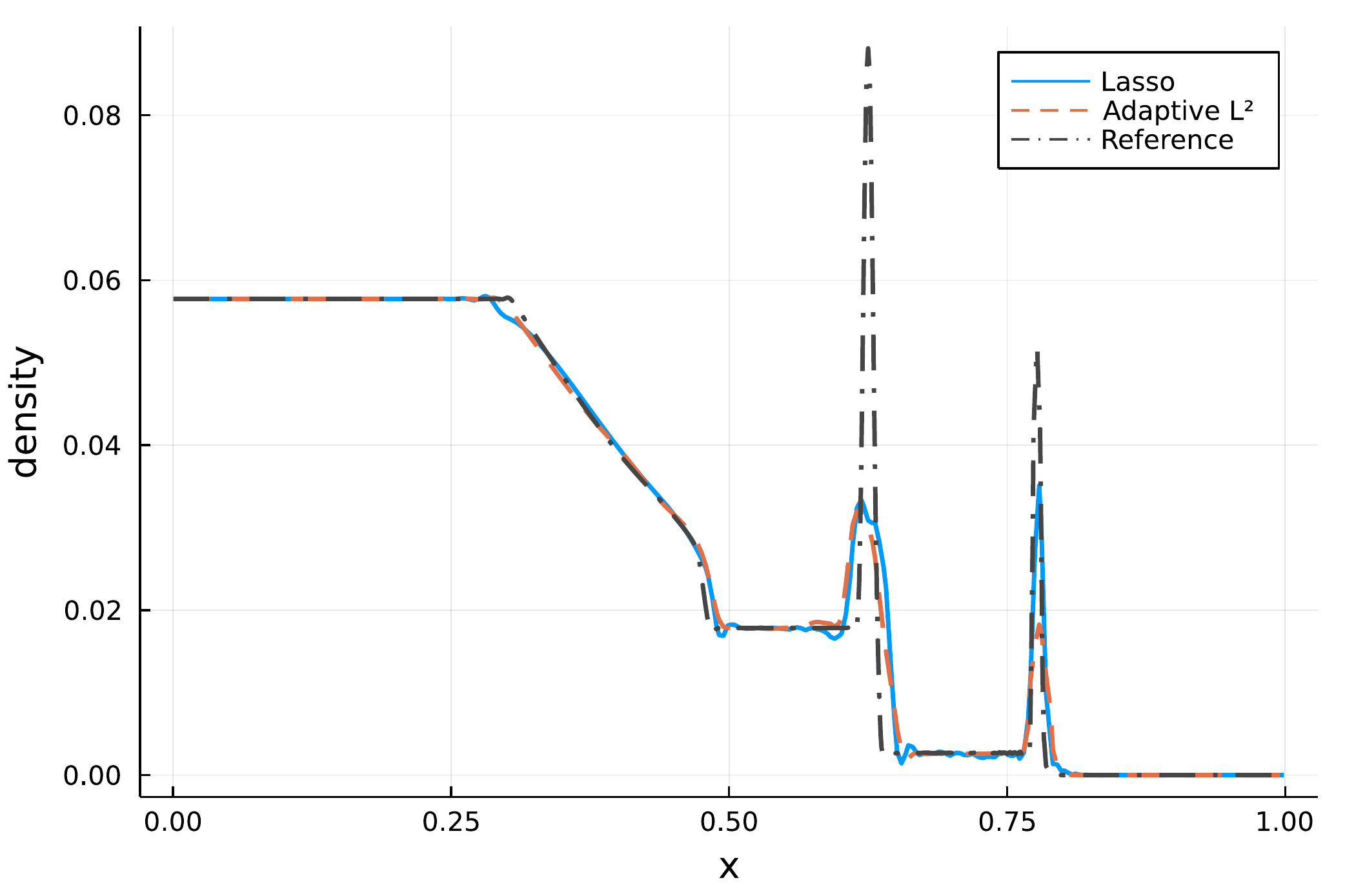}
    \end{minipage}
    \begin{minipage}{0.49\textwidth}
        \includegraphics[width=\textwidth]{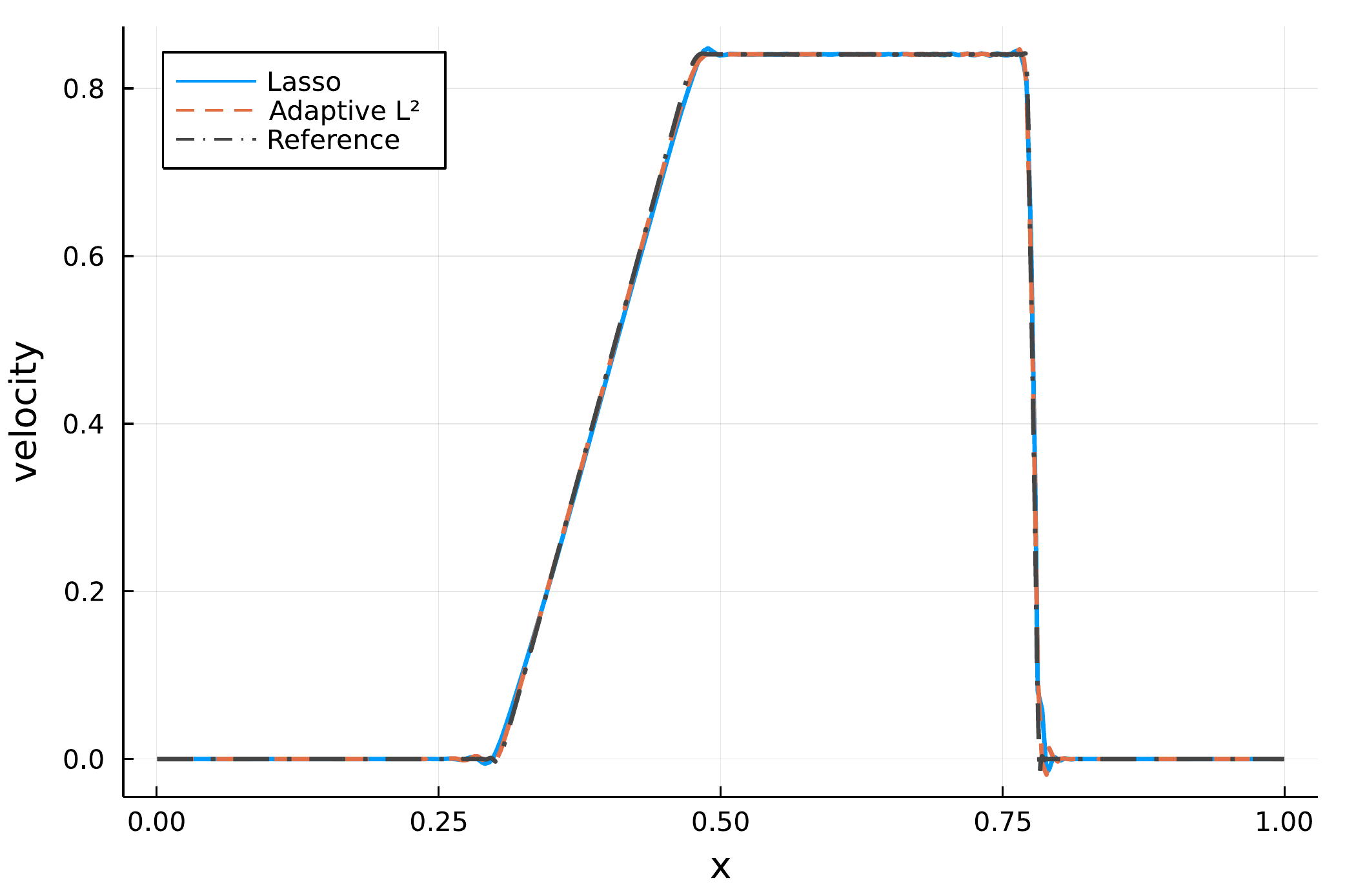}
    \end{minipage}
    \begin{minipage}{0.49\textwidth}
        \includegraphics[width=\textwidth]{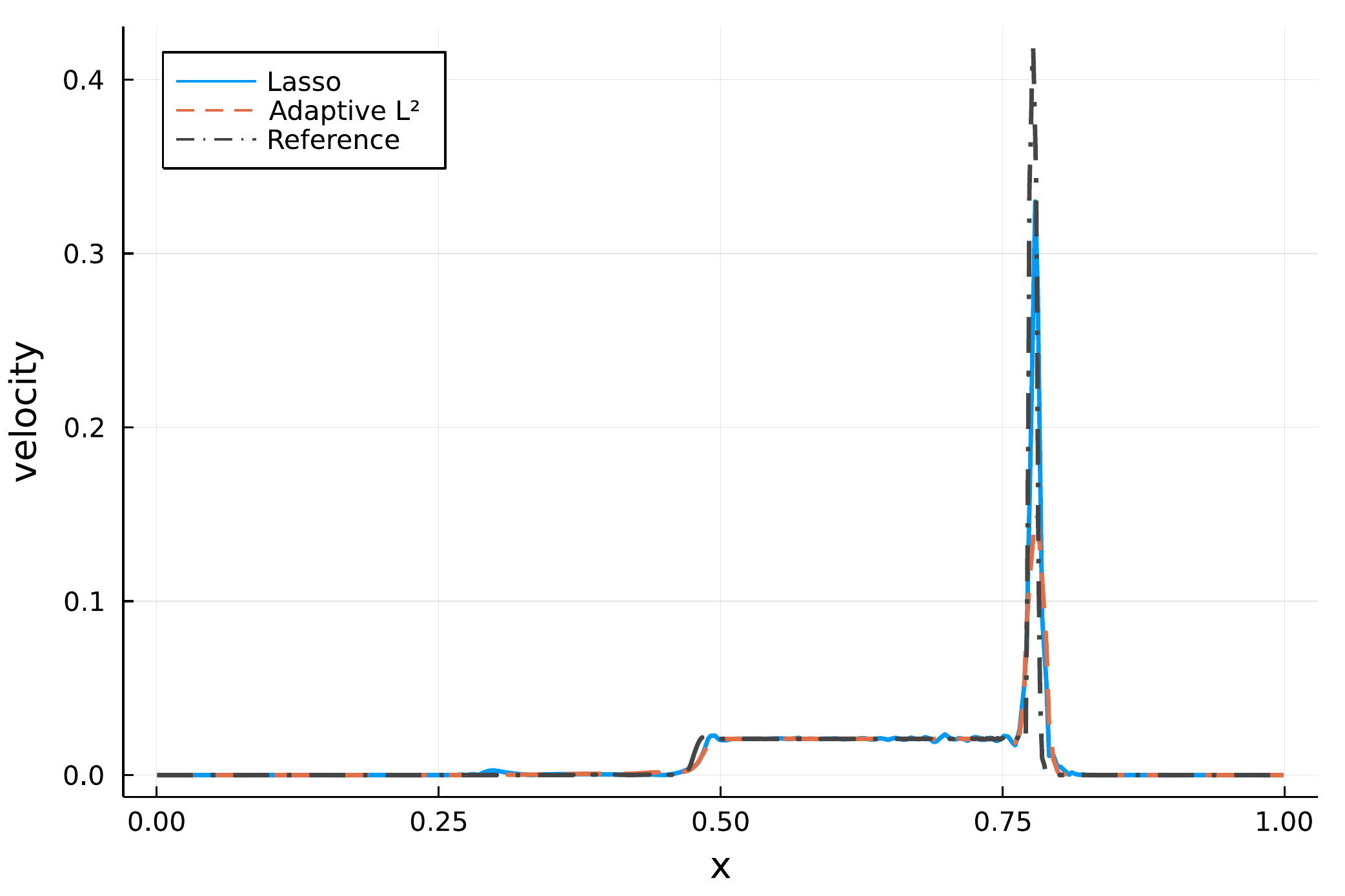}
    \end{minipage}
    \begin{minipage}{0.49\textwidth}
        \includegraphics[width=\textwidth]{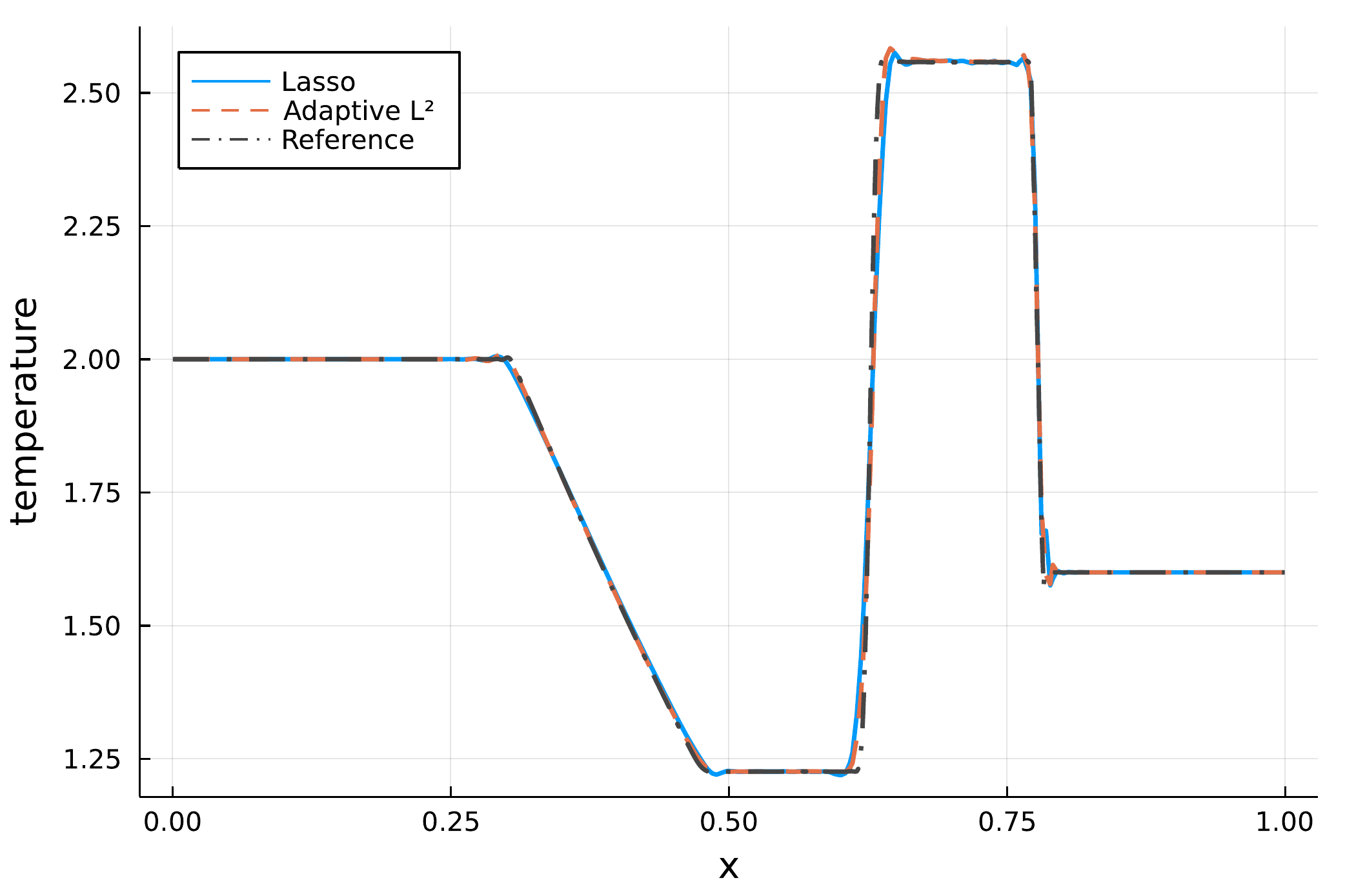}
    \end{minipage}
    \begin{minipage}{0.49\textwidth}
        \includegraphics[width=\textwidth]{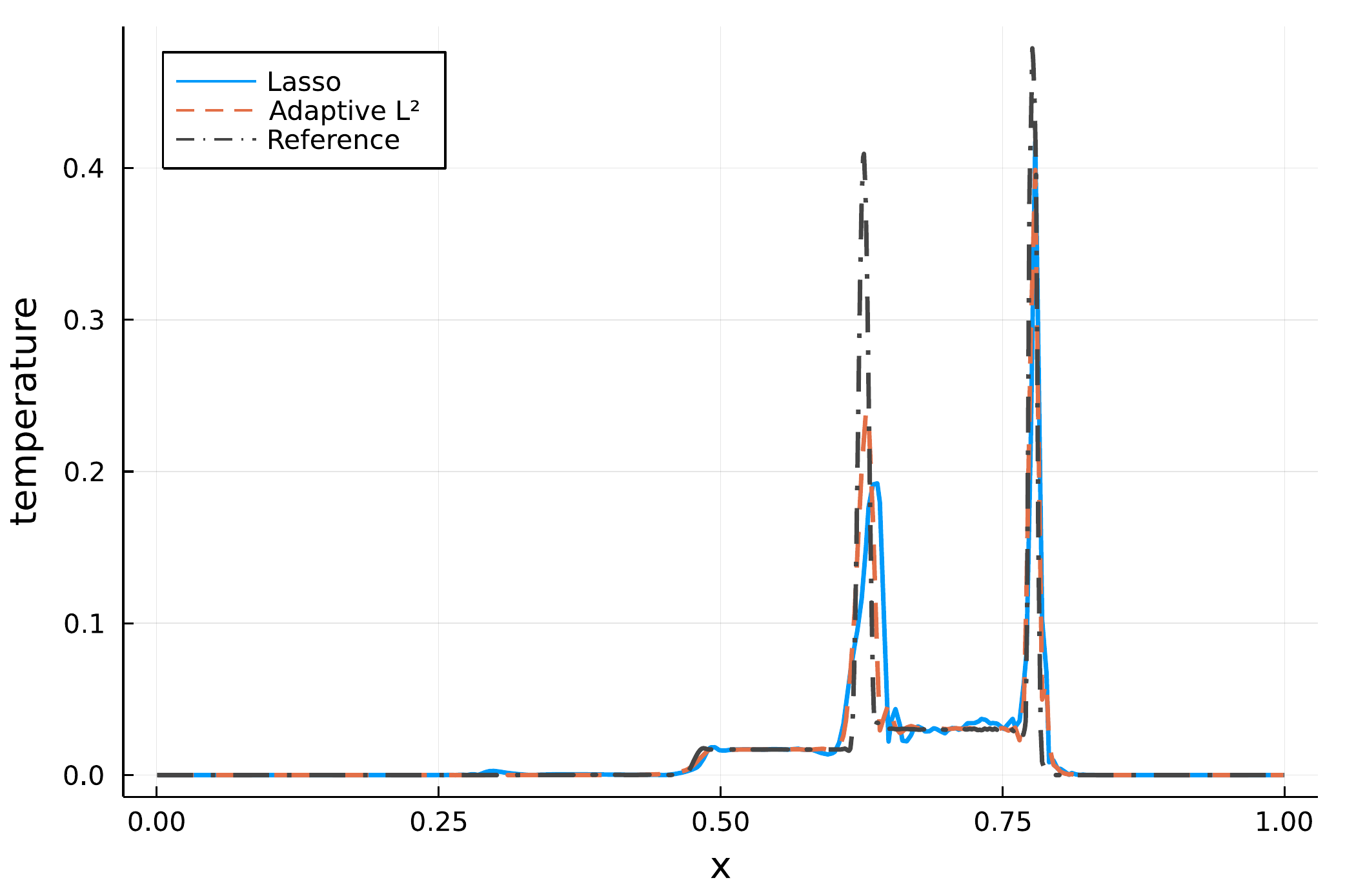}
    \end{minipage}
    \caption{The expected values (left column) and standard deviations (right column) of density, velocity and temperature in the Sod shock tube at $t=0.15$ under stochastic initial density.}
    \label{fig:sod1}
\end{figure}

The expected values and standard deviations of density, velocity and temperature inside the shock tube at $t=0.15$ are shown in Fig. \ref{fig:sod1} .
The collocation results produced by the second-order finite volume method \cite{xiao2021kinetic} with 500 elements are plotted as benchmark.
As can be seen, both filters robustly capture the expected structures of the rarefaction wave, the contact discontinuity and the shock wave.

\begin{figure}
    \centering
    \begin{minipage}{0.49\textwidth}
        \includegraphics[width=\textwidth]{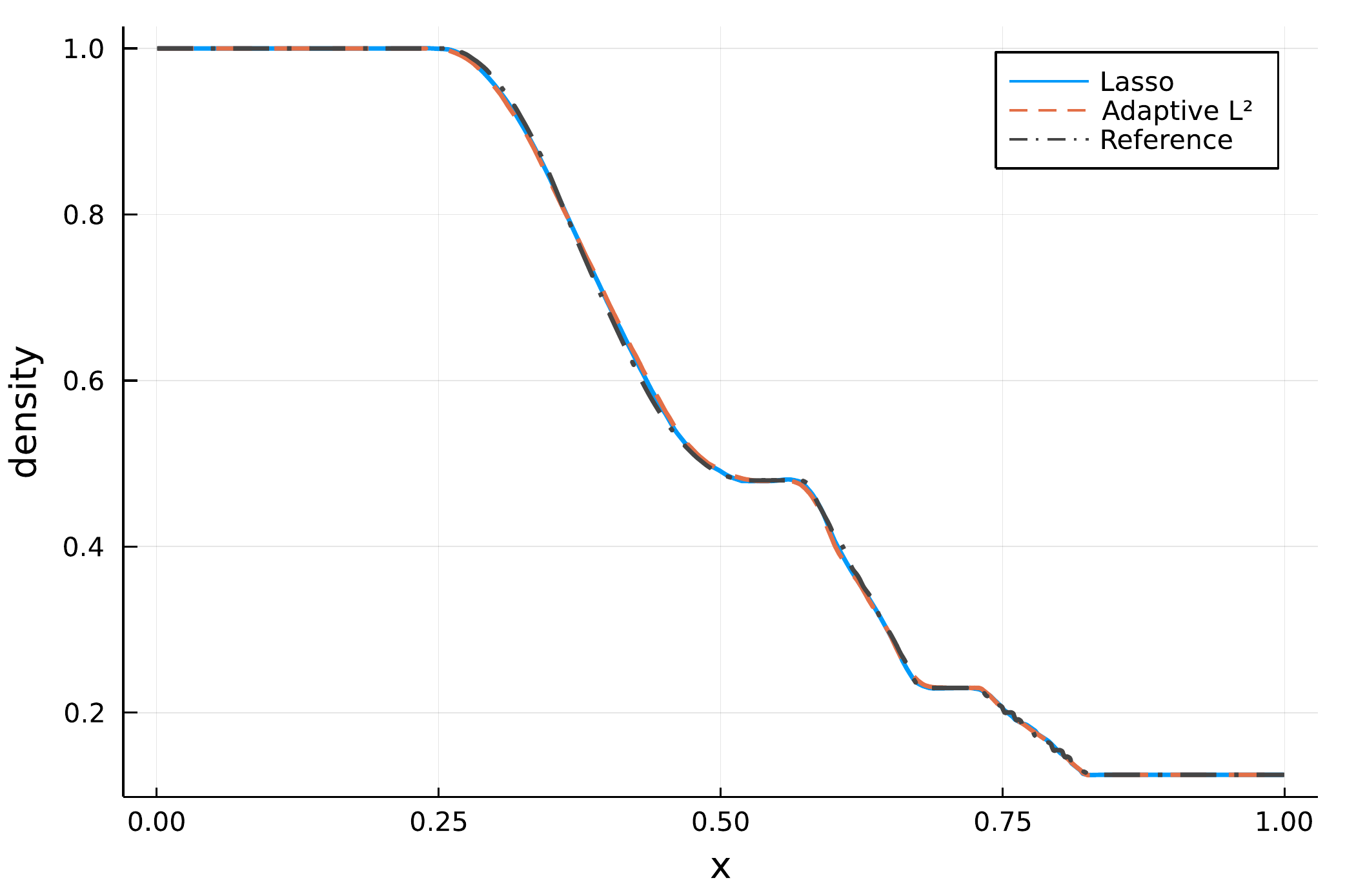}
    \end{minipage}
    \begin{minipage}{0.49\textwidth}
        \includegraphics[width=\textwidth]{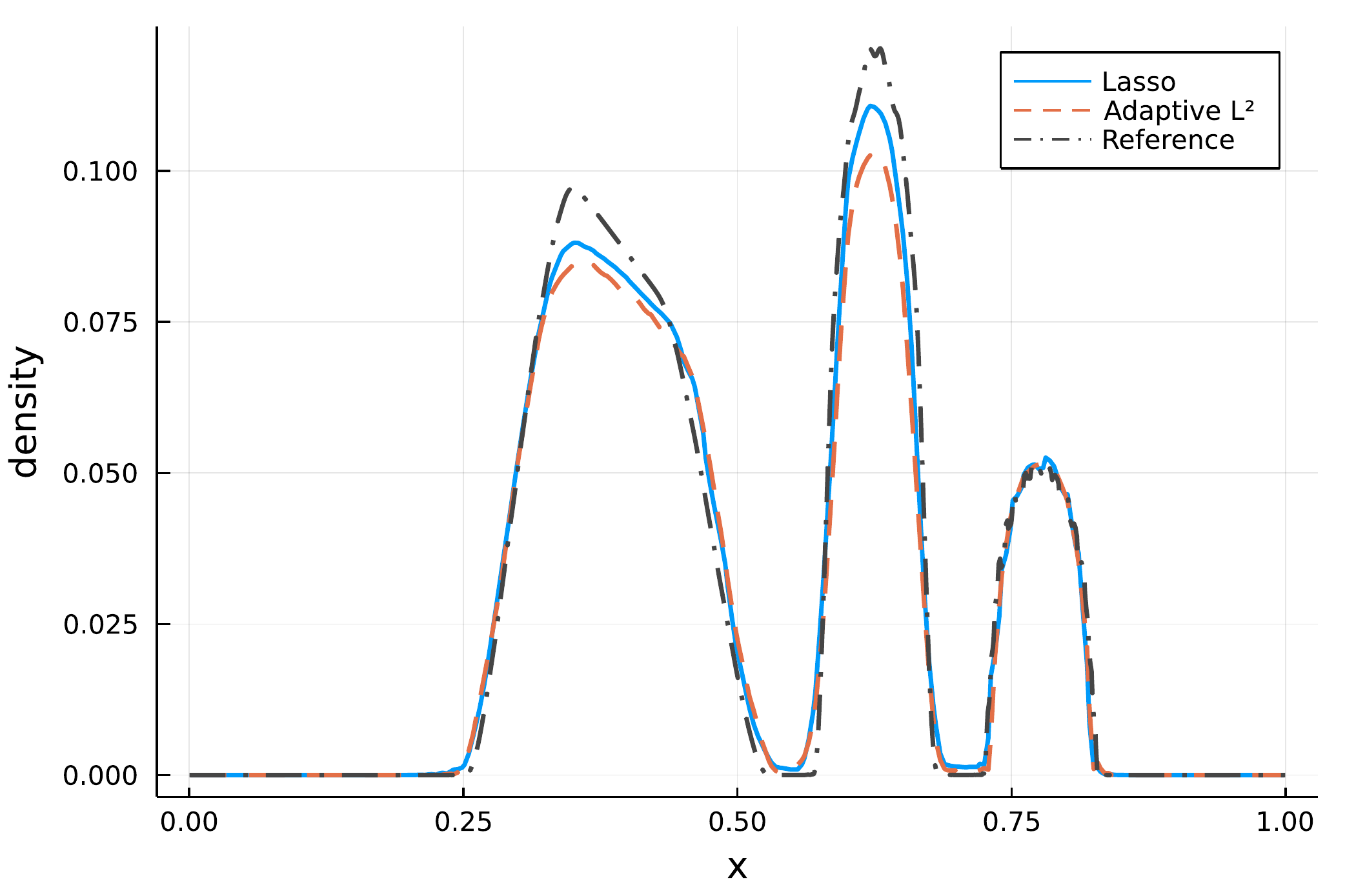}
    \end{minipage}
    \begin{minipage}{0.49\textwidth}
        \includegraphics[width=\textwidth]{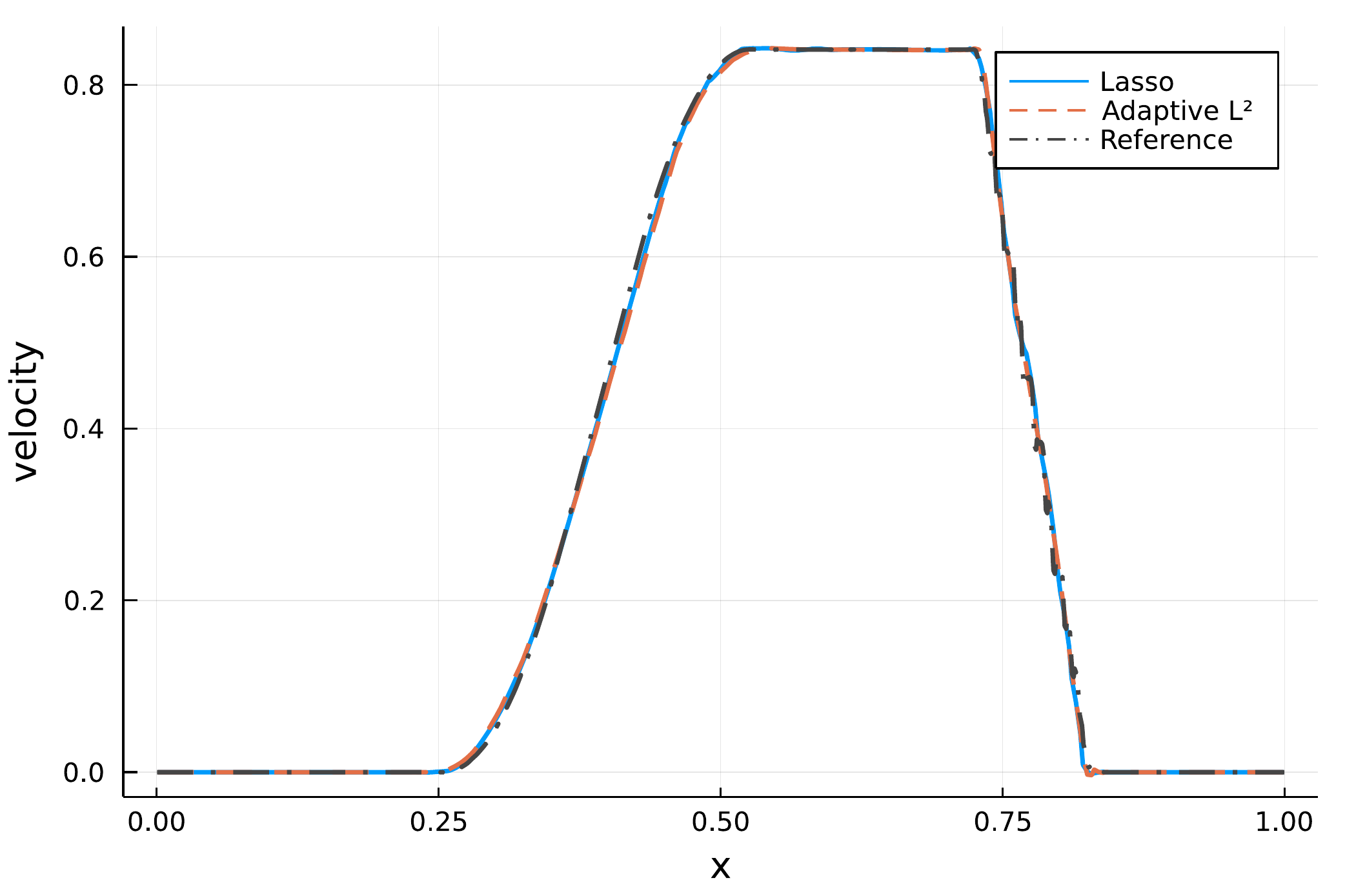}
    \end{minipage}
    \begin{minipage}{0.49\textwidth}
        \includegraphics[width=\textwidth]{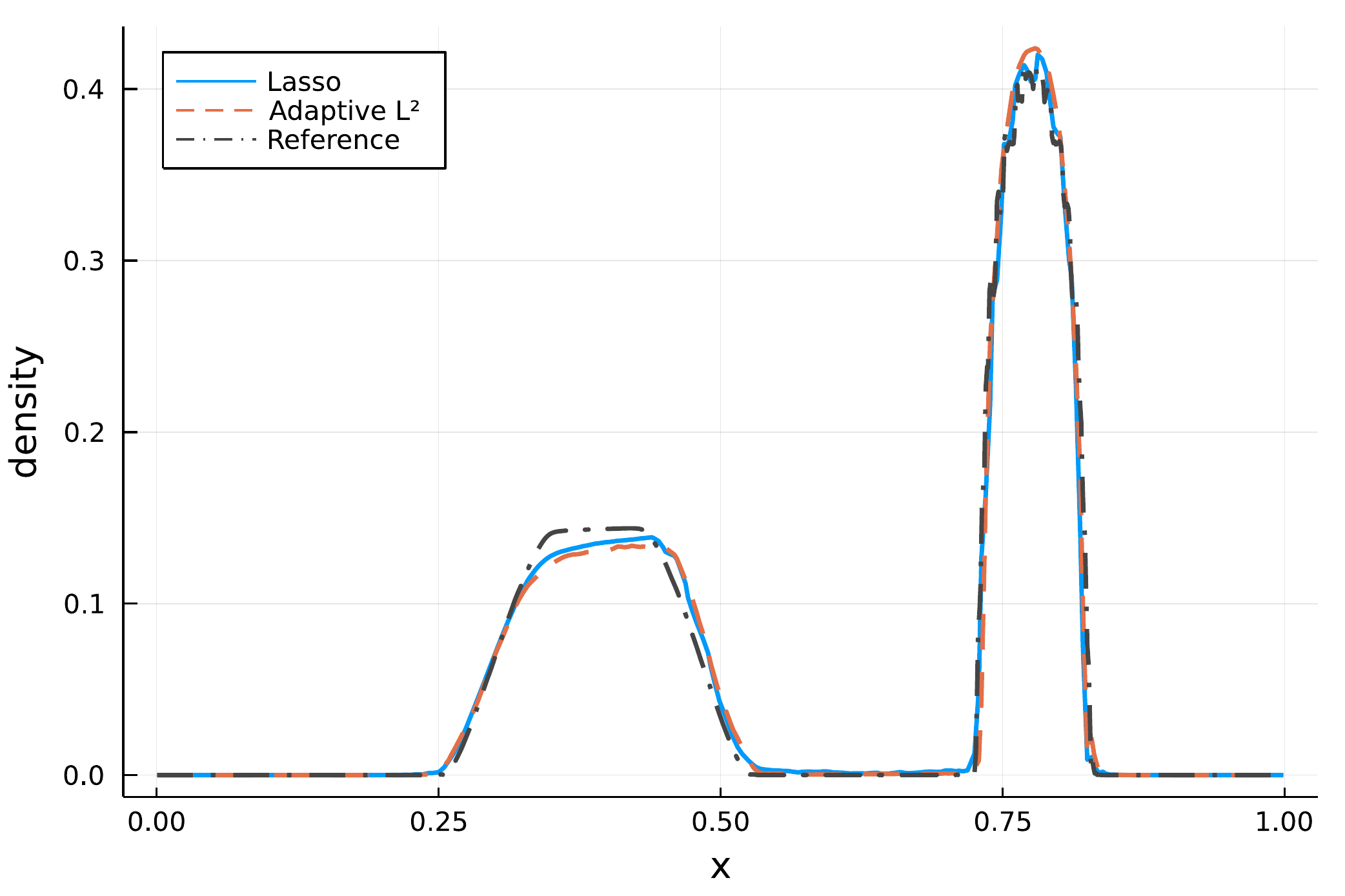}
    \end{minipage}
    \begin{minipage}{0.49\textwidth}
        \includegraphics[width=\textwidth]{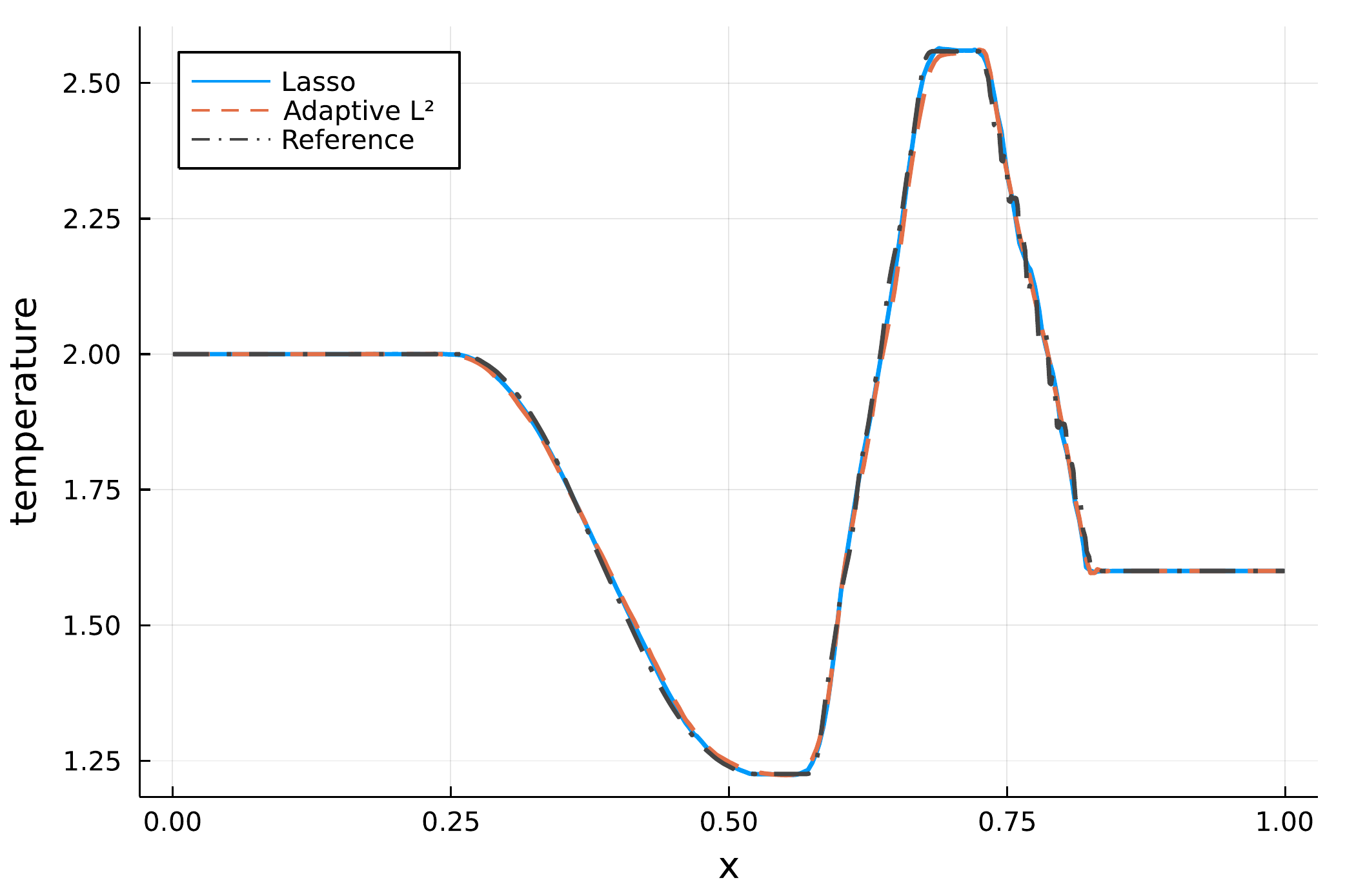}
    \end{minipage}
    \begin{minipage}{0.49\textwidth}
        \includegraphics[width=\textwidth]{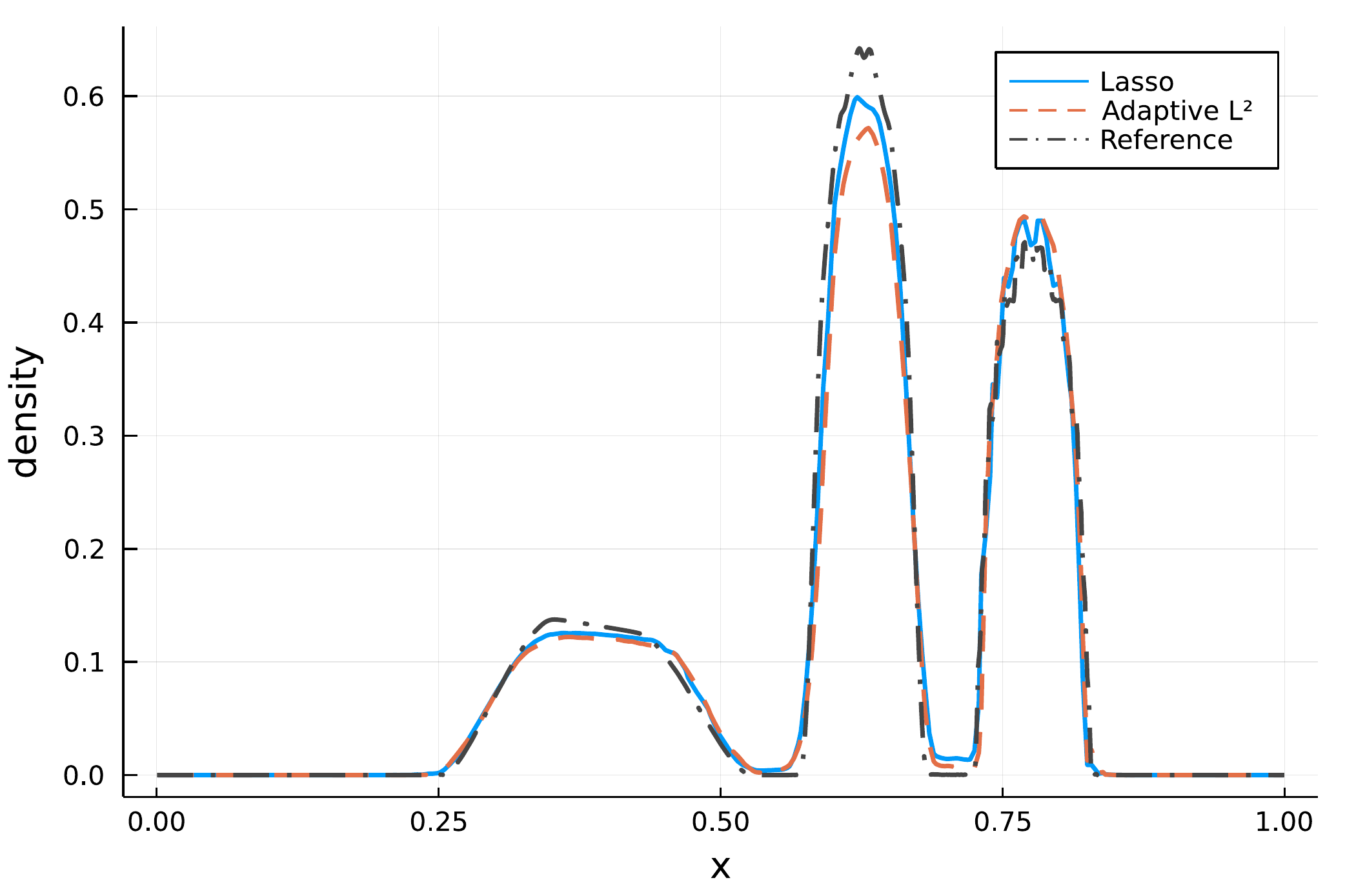}
    \end{minipage}
    \caption{The expected values (left column) and standard deviations (right column) of density, velocity and temperature in the Sod shock tube at $t=0.15$ under stochastic initial discontinuity location.}
    \label{fig:sod2}
\end{figure}

In the second case, the standard SG scheme fails within the beginning iterations due to the strong discontinuity in random space.
The filters together with the positivity-preserving limiter play a good role in mitigating the oscillations and enabling the simulation.
Similar as for the Burgers' equation, the Lasso filter presents less dissipation in the random space and thus results in sharper standard deviation values.
The slight oscillations around the shock wave can be further dampened by the adaptive $L^2$ filter, as shown in Fig. \ref{fig:sod2}.

\subsection{Shock-vortex interaction}

In the last case let us turn to the two-dimensional Euler equations,
\begin{equation}
    \frac{\partial}{\partial t}\left(\begin{array}{c}
    \rho \\
    \rho U \\
    \rho V \\
    \rho E
    \end{array}\right)
    +\frac{\partial}{\partial x} \left(\begin{array}{c}
    \rho {U} \\
    \rho {U}^2 +p \\
    \rho U V \\
    (\rho E + p)U
    \end{array}\right)
    +\frac{\partial}{\partial y} \left(\begin{array}{c}
    \rho {U} \\
    \rho UV \\
    \rho V^2 + p \\
    (\rho E + p)V
    \end{array}\right)
    =\left(\begin{array}{c}
    0 \\
    0 \\
    0 \\
    0
    \end{array}\right).
\end{equation}
We consider the shock-vortex interaction problem,
where the longitudinal and transverse processes coexist in the flow domain under stochastic Mach numbers.
The right-propagating shock wave is initialized by the Rankine-Hugoniot condition,
\begin{equation}
\begin{aligned}
    &\rho_R=1, \ \rho_{L}=\frac{(\gamma+1) \mathrm{Ma}^{2}}{(\gamma-1) \mathrm{Ma}^{2}+2} \rho_R, \\
    &U_R=0, \ U_{L}=c\mathrm{Ma} - \frac{(\gamma-1) \mathrm{Ma}^{2}+2}{(\gamma+1) \mathrm{Ma}^{2}}, \\
    &V_R=0, \ V_L = 0,\\
    &T_R=1, \ T_{L}=\frac{\left((\gamma-1) \mathrm{Ma}^{2}+2\right)\left(2 \gamma \mathrm{Ma}^{2}-\gamma+1\right)}{(\gamma+1)^{2} \mathrm{Ma}^{2}} T_{R},
\end{aligned}
\end{equation}
where the variables marked with $R$ and $L$ denote the upstream and downstream conditions, respectively.
The specific heat ratio is denoted by $\gamma$ and Ma is the Mach number.
The vortex is defined as an isentropic perturbation to the background fluid, 
\begin{equation}
\begin{aligned}
    &(\delta U, \ \delta V)=\zeta \eta e^{\mu\left(1-\eta^{2}\right)}(\sin \theta,-\cos \theta), \\
    &\delta T=-\frac{(\gamma-1) \zeta^{2}}{4 \mu \gamma} e^{2 \mu\left(1-\eta^{2}\right)}, \ \delta S=0,
\end{aligned}
\end{equation}
where $S=\ln (p/\rho^\gamma)$ is the entropy.
A polar coordinate $(r,\theta)$ is formulated by the center of the vortex $(x_c,y_c)$, where the radius is given by $r=\sqrt{(x-x_c)^2+(y-y_c)^2}$ and $\eta = r/r_c$.
The parameter $\kappa$ defines the strength of the vortex, $\mu$ indicates the decay rate of the vortex, and $r_c$ is the critical radius at which the vortex holds the maximum strength.
The initial flow field is therefore set as,
\begin{equation}
\begin{aligned}
    & \mathbf v(t=0,x,y,z):= \left[\begin{array}{c}
    \rho \\
    U \\
    V \\
    p 
    \end{array}\right] = \begin{cases} \mathbf v_{L} + \delta \mathbf v, & x<x_s, \\ \mathbf v_{R} + \delta \mathbf v, & x \geq x_s ,\end{cases},
\end{aligned}
\end{equation}
where $x_s$ is the location of the shock.
The detailed computational setup can be found in Table \ref{tab:vortex}. 
\begin{table}[htbp]
	\caption{Computational setup of shock-vortex interaction problem.} 
	\centering
	\begin{tabular}{lllllllll} 
		\hline
		$t$ & $x$ & $y$ & $z$ & $N_x$ & $N_y$ \\
		$(0,1]$ & $[0,2]$ & $[0,1]$ & $[-1,1]$ & $100$ & 50 \\ 
		\hline
		Points & $N_p$ & Correction & Ma & $x_s$ & $x_c$ \\
		Legendre & 3 & Radau & $\mathcal U(1.06,1.18)$ & 0.25 & 0.8 \\
		\hline
		$y_c$ & $r_c$ & $\zeta$ & $\mu$ & gPC & $N_c$ \\
		0.5 & 0.05 & 0.25 & 0.204 & Legendre & 5 \\
		\hline
		$N_q$ & Flux & Integrator & Boundary & CFL & Filter \\
		9 & HLL & Bogacki-Shampine & Reflection & 0.1 & $L^2$ \\
		\hline
		$s_0$ & $\kappa$ & $\varepsilon_1$ & $\varepsilon_2$ \\
		$-3 \log 2$ & 4 & 0.6 & 1 \\
		\hline
	\end{tabular} 
	\label{tab:vortex}
\end{table}

\begin{figure}
    \centering
    \subfigure[Expectation]{
		\includegraphics[width=0.47\textwidth]{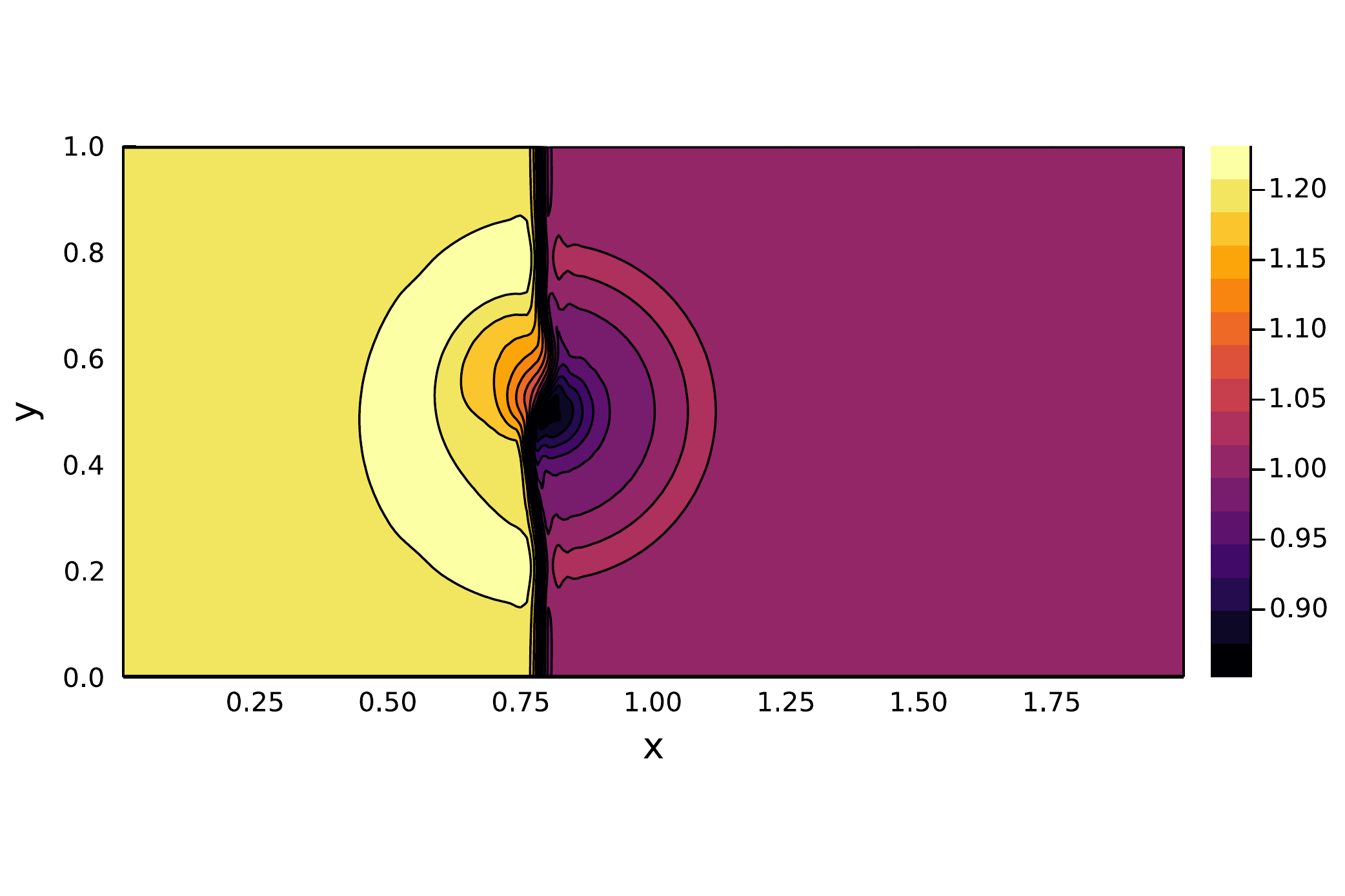}
	}
	\subfigure[Standard deviation]{
		\includegraphics[width=0.47\textwidth]{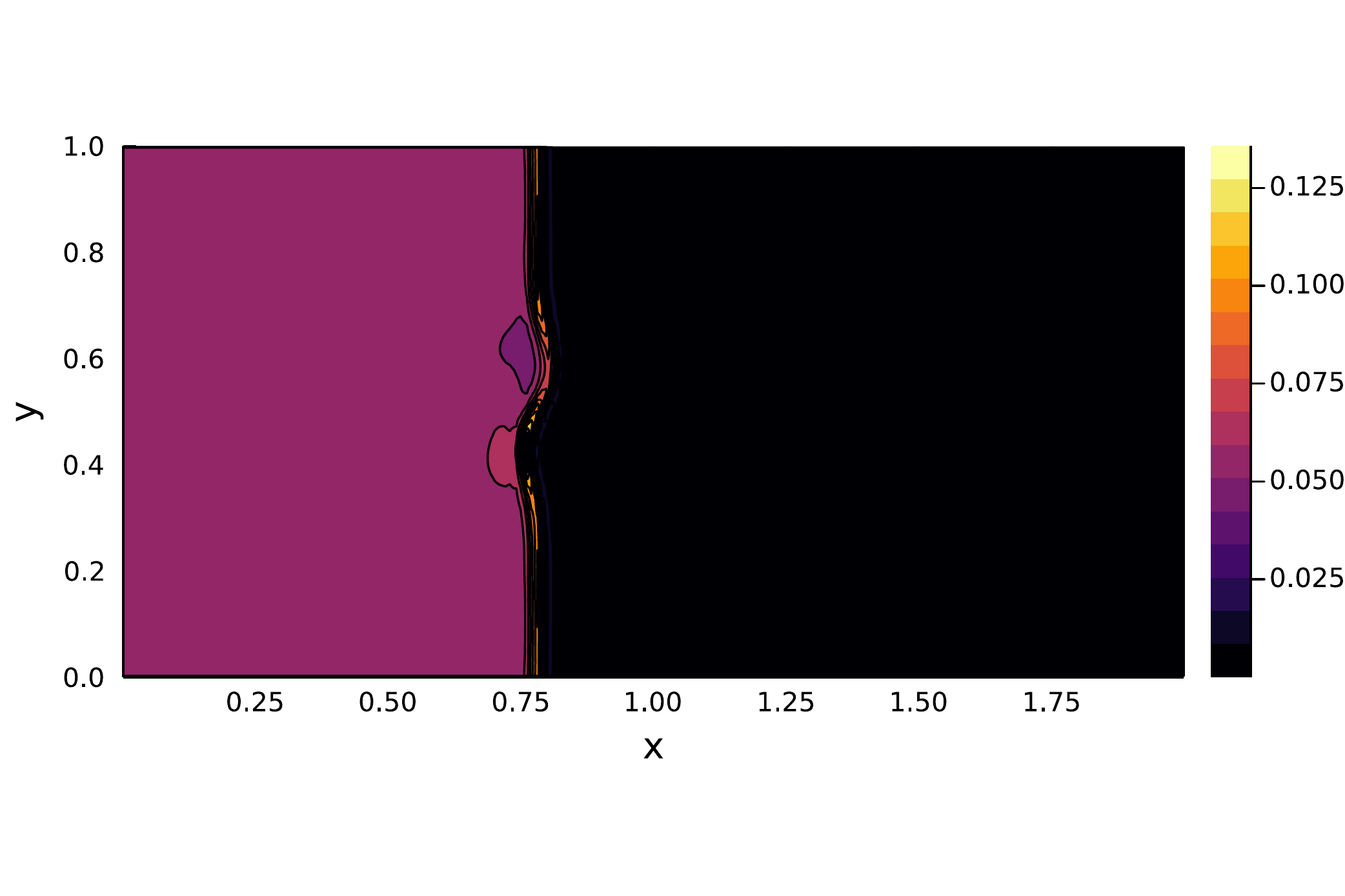}
	}
    \caption{Expected value and standard deviation of density in the shock-vortex interaction problem at $t=0.3$.}
    \label{fig:sv t1}
\end{figure}

\begin{figure}
    \centering
    \subfigure[Expectation]{
		\includegraphics[width=0.47\textwidth]{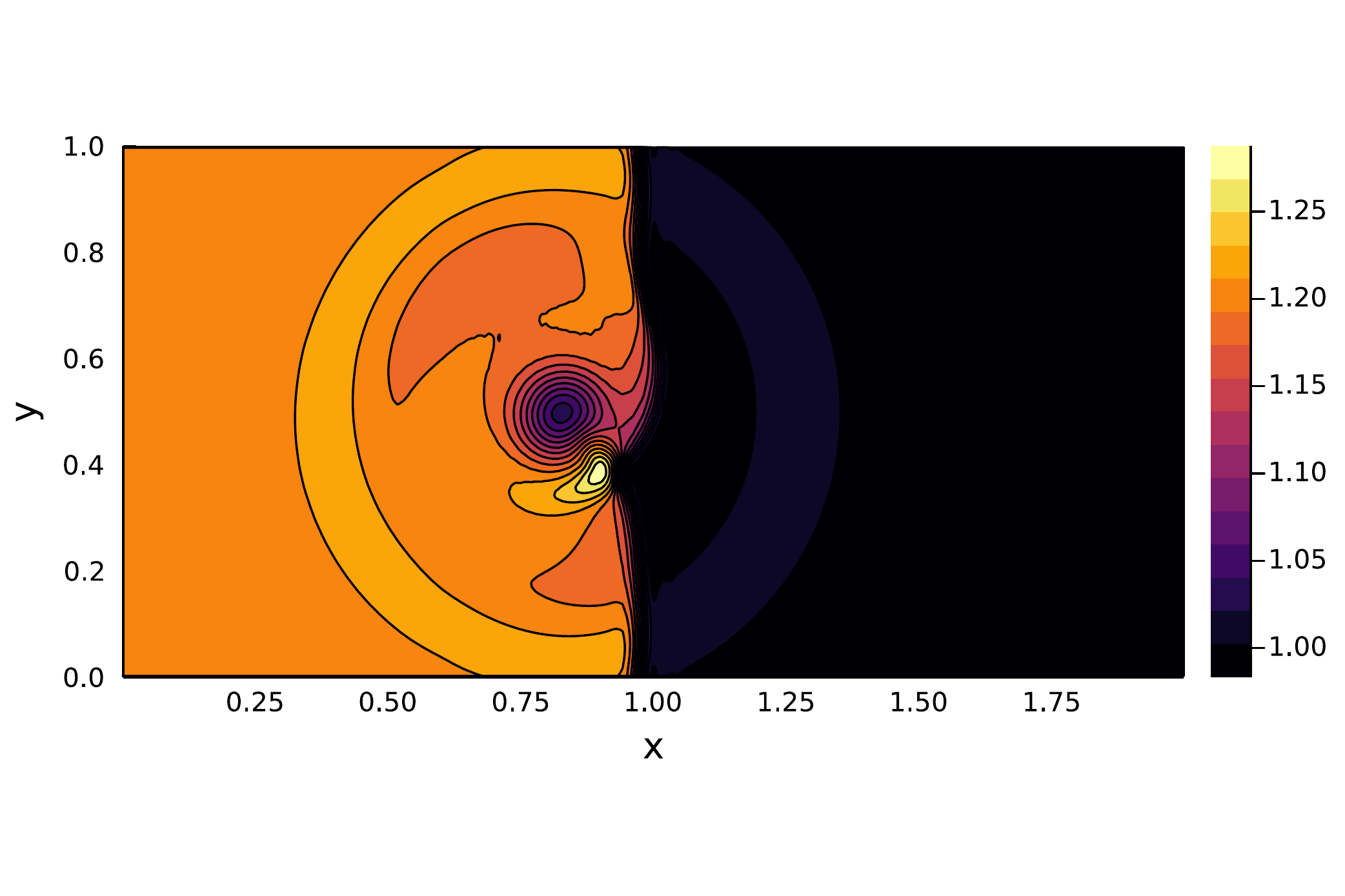}
	}
	\subfigure[Standard deviation]{
		\includegraphics[width=0.47\textwidth]{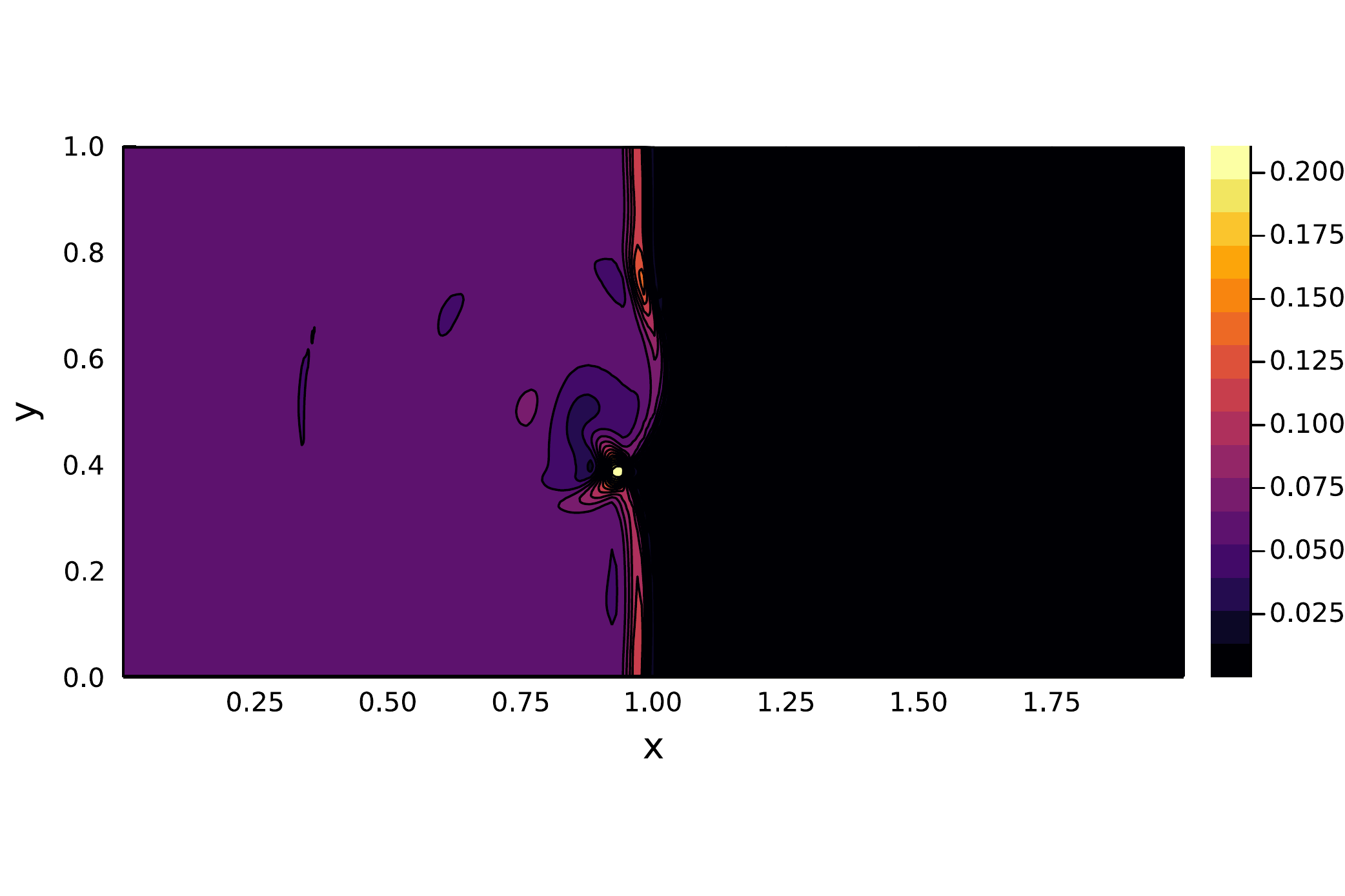}
	}
    \caption{Expected value and standard deviation of density in the shock-vortex interaction problem at $t=0.5$.}
    \label{fig:sv t2}
\end{figure}

\begin{figure}
    \centering
    \subfigure[Expectation]{
		\includegraphics[width=0.47\textwidth]{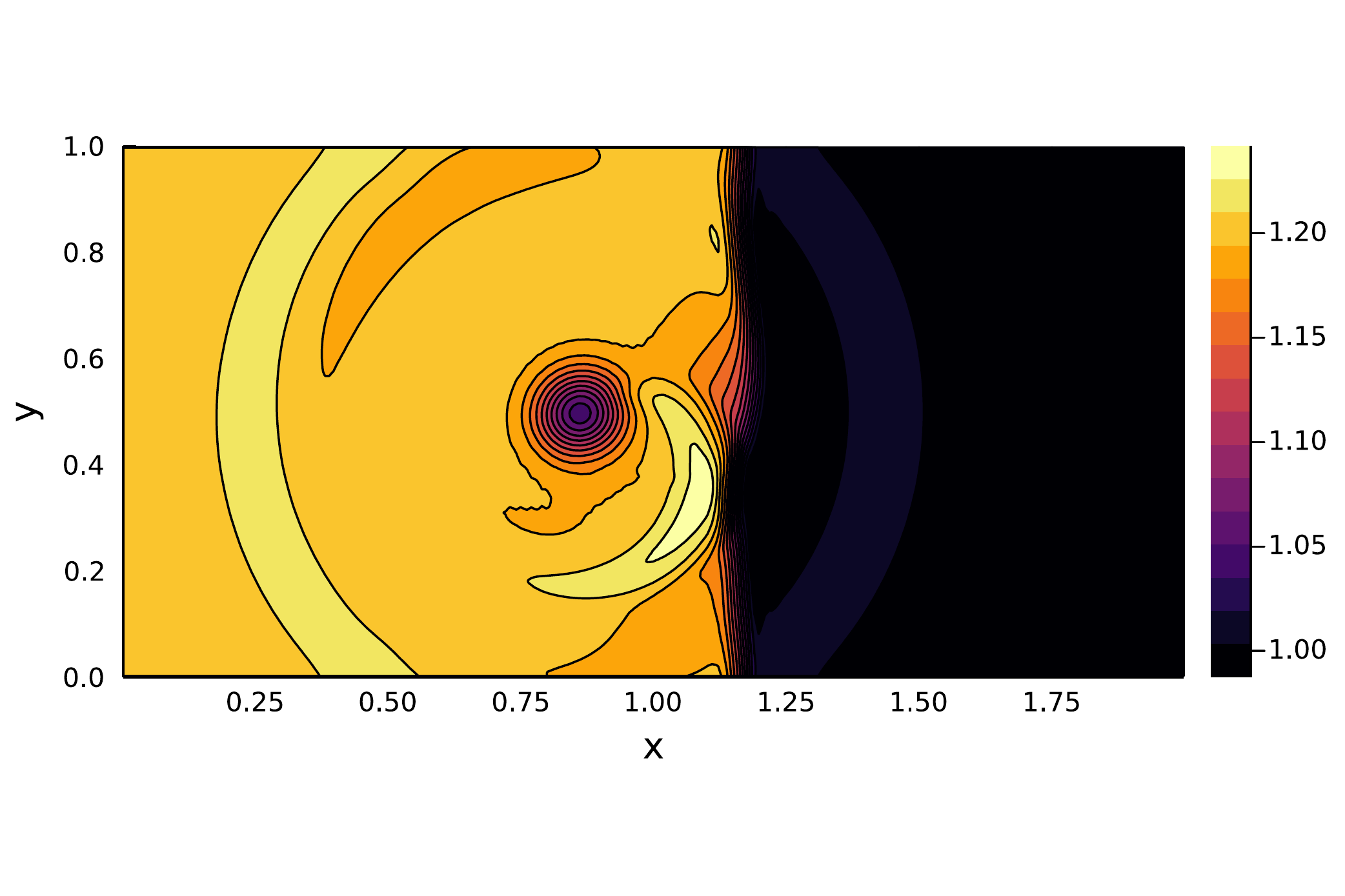}
	}
	\subfigure[Standard deviation]{
		\includegraphics[width=0.47\textwidth]{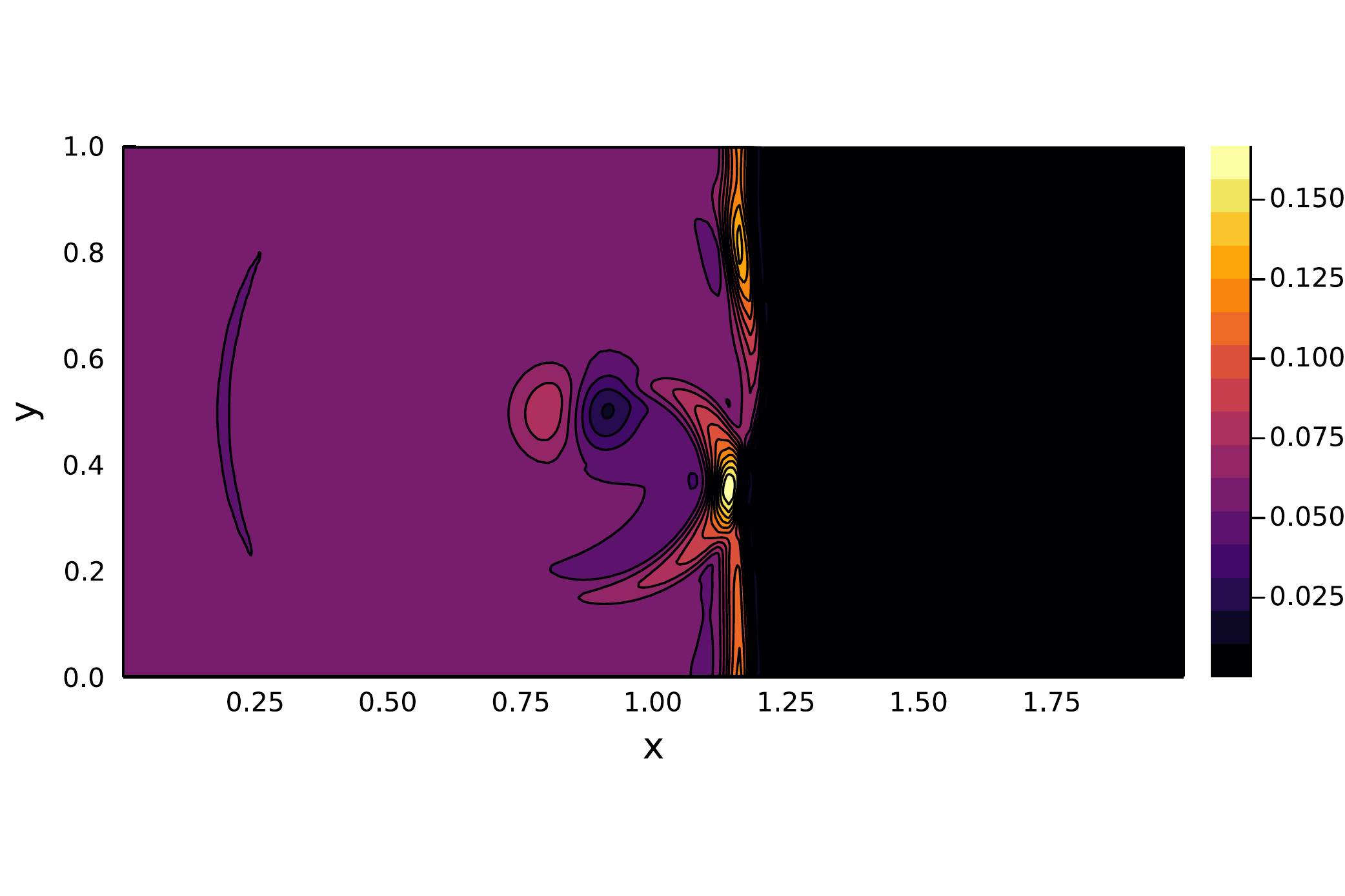}
	}
    \caption{Expected value and standard deviation of density in the shock-vortex interaction problem at $t=0.7$.}
    \label{fig:sv t3}
\end{figure}

\begin{figure}
    \centering
    \subfigure[Expectation]{
		\includegraphics[width=0.47\textwidth]{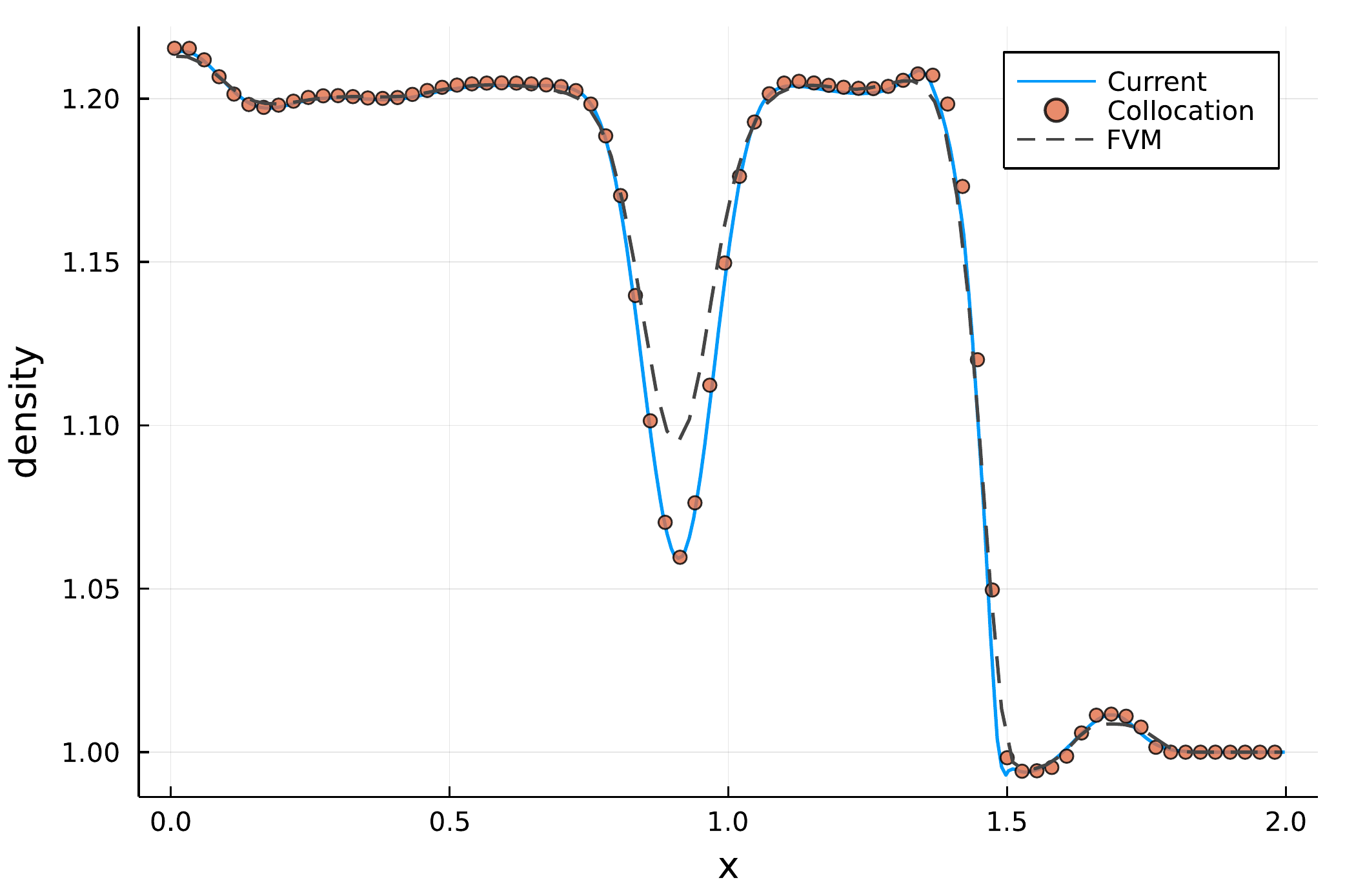}
	}
	\subfigure[Standard deviation]{
		\includegraphics[width=0.47\textwidth]{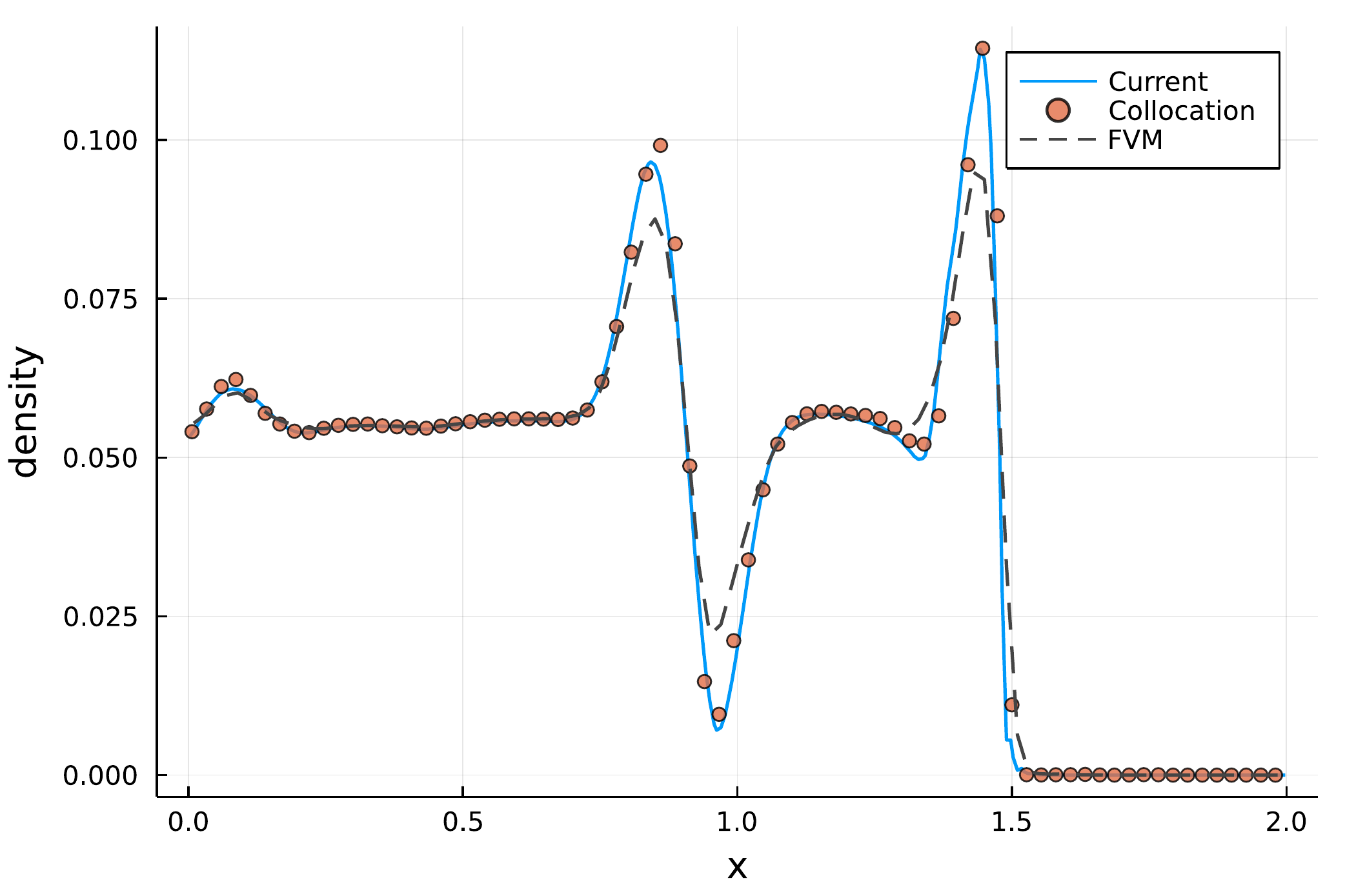}
	}
    \caption{Expected value and standard deviation of density in the shock-vortex interaction problem at $t=1$.}
    \label{fig:sv t4 density}
\end{figure}

\begin{figure}
    \centering
    \subfigure[Expectation]{
		\includegraphics[width=0.47\textwidth]{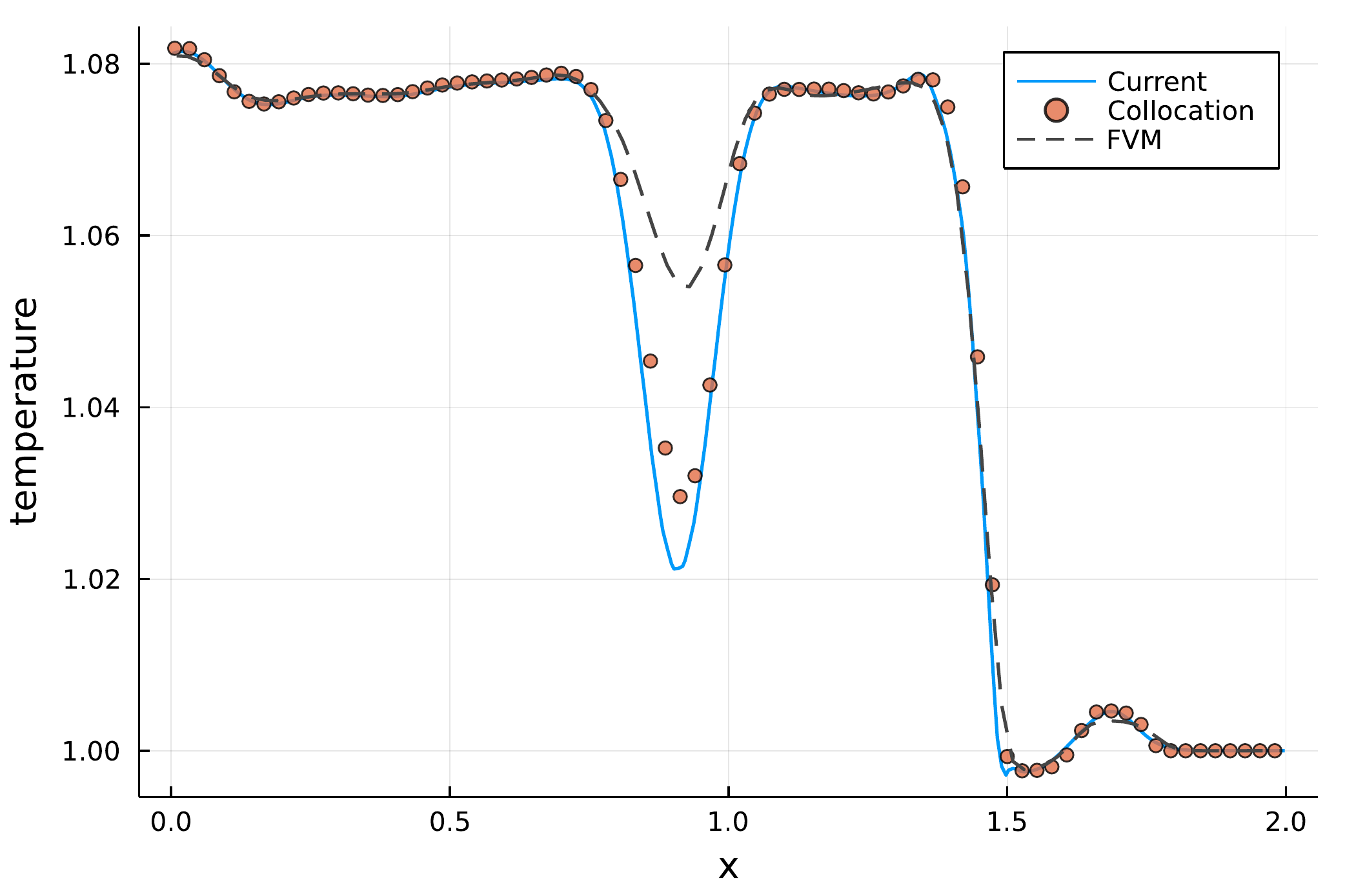}
	}
	\subfigure[Standard deviation]{
		\includegraphics[width=0.47\textwidth]{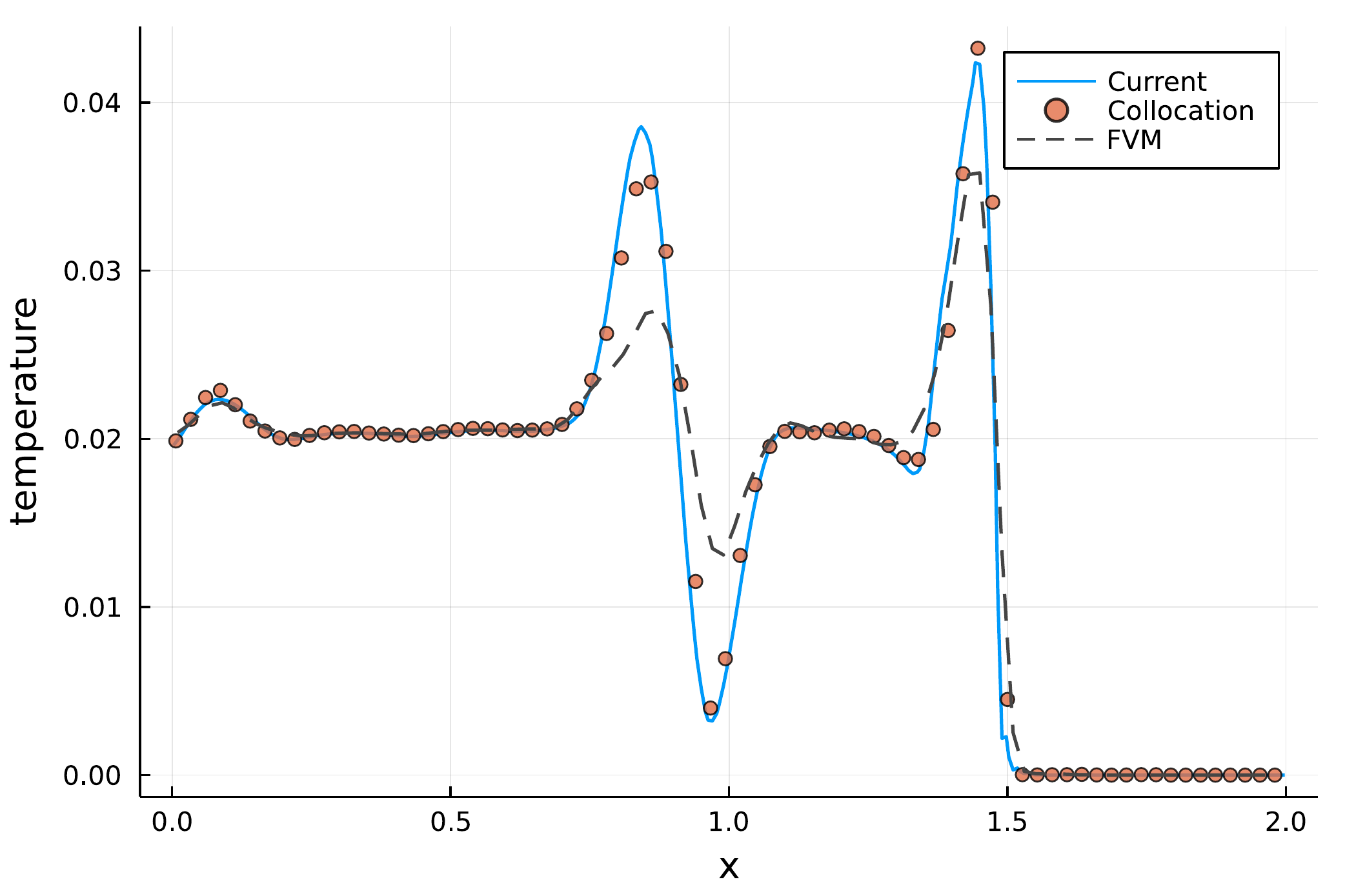}
	}
    \caption{Expected value and standard deviation of temperature in the shock-vortex interaction problem at $t=1$.}
    \label{fig:sv t4 temperature}
\end{figure}

Fig. \ref{fig:sv t1}, \ref{fig:sv t2} and \ref{fig:sv t3} present the expected values and standard deviations of density contours at $t=0.3$, $0.5$ and $0.7$. 
As shown, the fine structures emerging from the interaction between longitudinal and transverse fluid processes are robustly captured by the current scheme.
The role of shock and vortex as source terms of uncertainties is clearly demonstrated.
Fig. \ref{fig:sv t4 density} and \ref{fig:sv t4 temperature} provide the profiles of density and temperature along the horizontal central line.
The collocation results produced by the deterministic flux reconstruction method and the second-order finite volume method \cite{xiao2021kinetic} with the same amount of elements are plotted for comparison.
It is clear that the current Galerkin scheme provides results equivalent to the benchmark collocation solutions.
Benefiting from the higher-order interpolations, the accuracy and fidelity of solutions are greatly improved compared to the second-order finite volume results.


\section{Conclusion}

The development of higher-fidelity numerical schemes is crucial in computational fluid dynamics.
In this paper, we present the first flux reconstruction stochastic Galerkin method for the study of uncertainty propagation.
Benefiting from the uniform spectral discretization, an accurate approximation of solutions can be achieved, and the numerical behaviors of the scheme in spatial and random domain are consistent.
The nodal and modal representations can be transformed naturally based on orthogonal polynomials and solution collocation points.
A family of multi-dimensional filters are developed to mitigate the Gibbs phenomenon and a positivity-preserving limiter is employed to preserve physically realizable solutions.
As a result, the current scheme is able to solve cross-scale problems, where resolved and unresolved regions coexist in the flow domain.
It provides a powerful tool for the study of sensitivity analysis and uncertainty propagation, and the performance is demonstrated through numerical experiments.

For future work, it is possible to apply the scheme to other complex systems, e.g., astrophysics \cite{xiao2017well}, particle transports \cite{xiao2020velocity}, and plasma physics \cite{xiao2021stochastic}.
An alternative to a hyperbolicity-preserving limiter is the careful alteration of the SG system itself, such that its hyperbolicity domain is significantly enlarged, or possibly the whole space. This approach of deriving globally hyperbolic models has been successfully applied for kinetic equations and free-surface flows, see \cite{Cai2013b,Koellermeier2014a,Fan2016,Koellermeier2020g,Koellermeier2020a}. A similar approach might be used in SG models to avoid using bound-preserving limiters in future work.

\appendix
\section{Parameter choice for exponential filter}
\label{app}

While the Lasso filter does not require numerical parameter choices, the exponential filter from section \ref{sec:exfilter} uses several parameters which need to be determined in applications. 

Different strategies exist in the literature.
In \cite{Hou2007} the filter parameter is chosen as $\alpha= 36$, together with the filter exponent $s=36$ to ensure that the last mode is damped to zero up to machine precision. However, the effect on the solution behavior is not clarified.
In \cite{Koellermeier2020a} the parameter choice was motivated with a number of heuristics. Firstly, the effect of the filter on the oscillation of the solution was investigated. Not surprisingly, it was found that larger parameters $\alpha$ smooth the solution and eventually recover positivity of the filtered distribution function. Secondly, a linear stability analysis of the model linearised around its equilibrium state revealed the damping factors for each mode. It was shown that the choice $\alpha = 36$ leads to small damping (i.e. less added diffusion) of the solution, while completely damping out the fastest mode. Lastly, the filter was tested with different parameters for the full model and the value $\alpha = 36$ indeed performed best with respect to the solution quality. While the best choice might depend on the size of the model, the choice of $\alpha = 36$ was robust in the test cases computed in \cite{Koellermeier2020a} and this value was therefore used for all further tests computed therein. 

\begin{figure}
    \centering
    \subfigure[Expectation]{
		\includegraphics[width=0.47\textwidth]{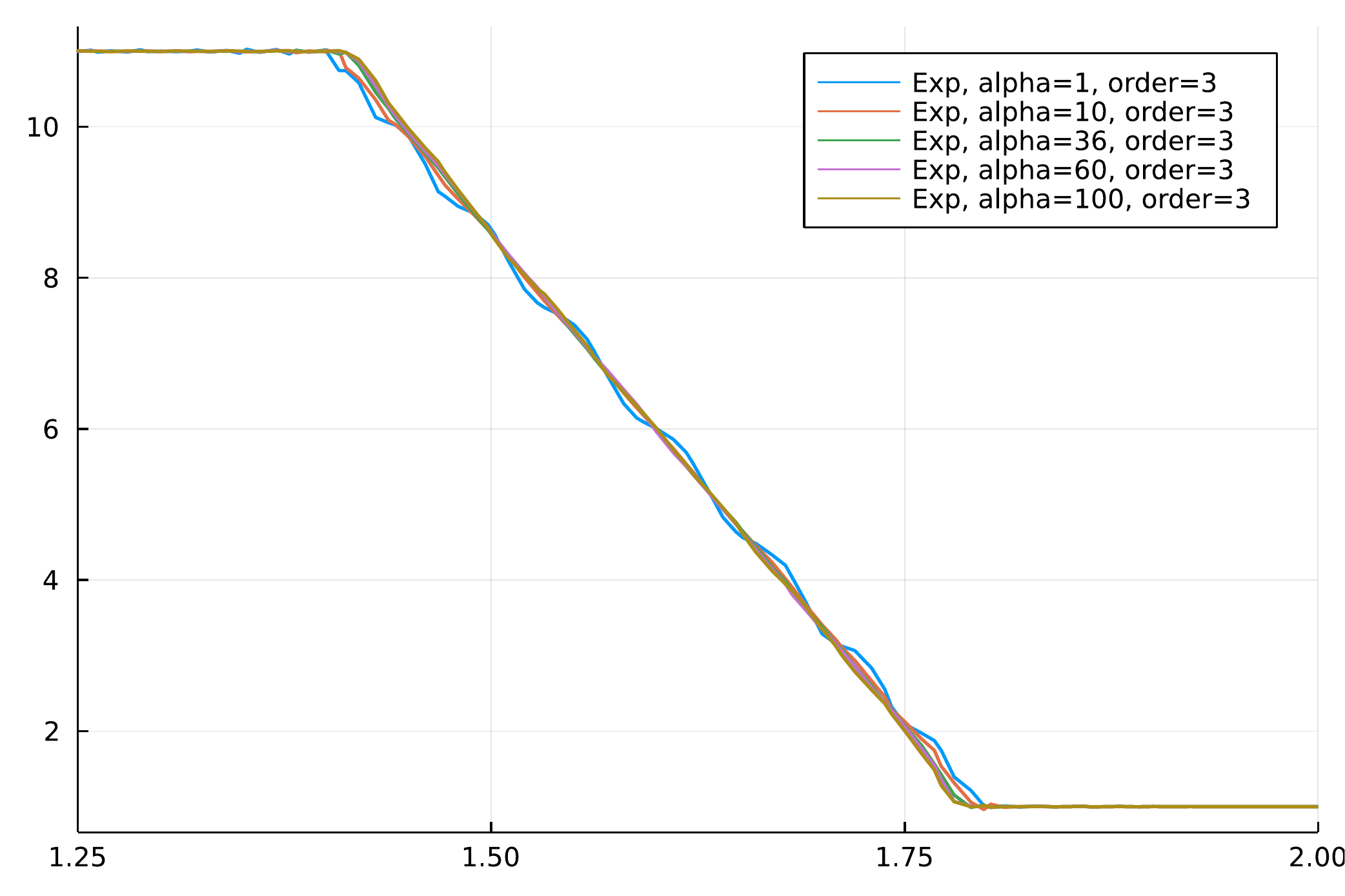}
	}
	\subfigure[Standard deviation]{
		\includegraphics[width=0.47\textwidth]{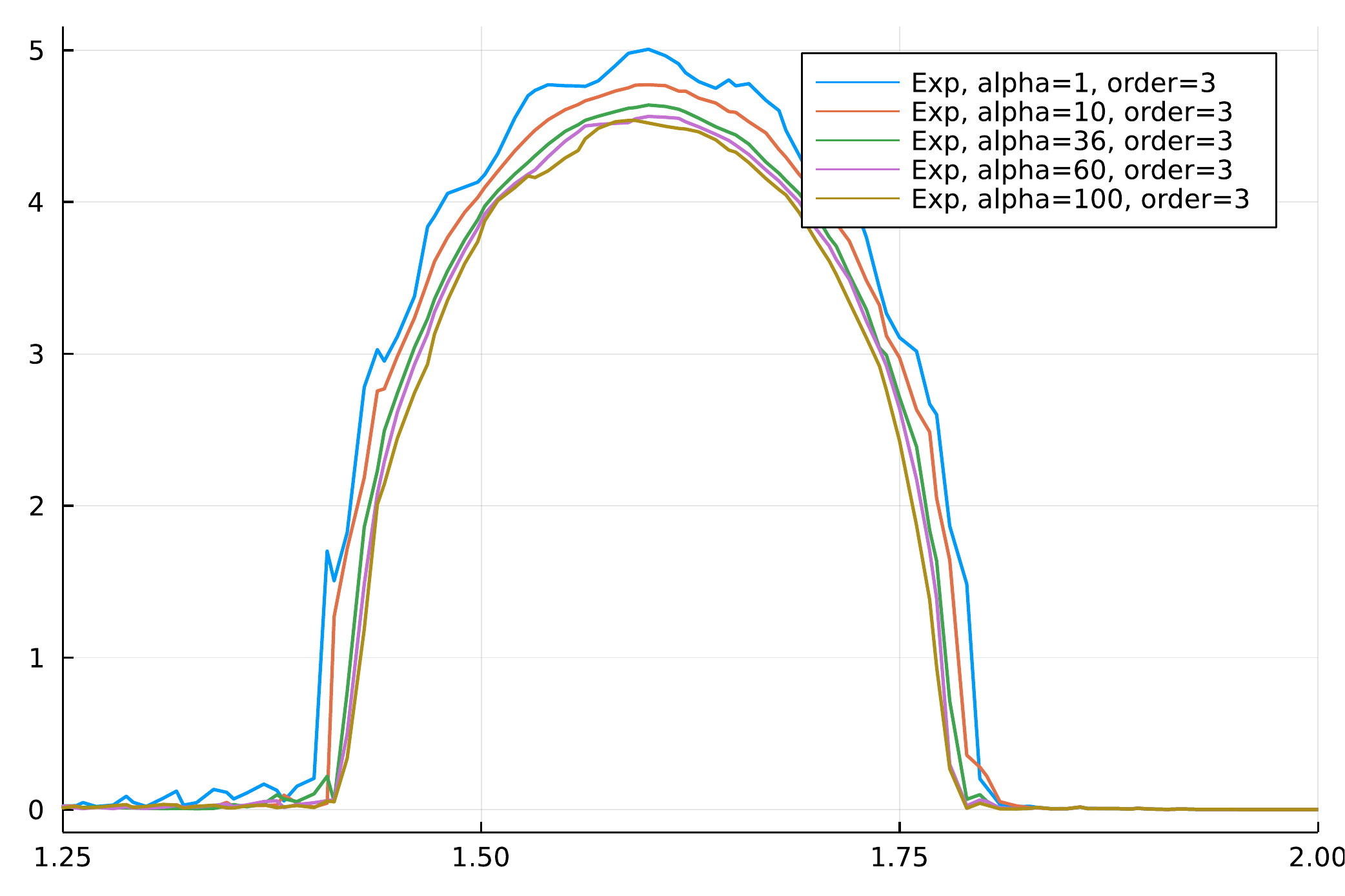}
	}
    \caption{Expected value and standard deviation for Burger's equation and varying filter parameters $\alpha$ of the exponential filter. The filter exponent is kept fixed at $s=3$ and we choose $N_* = 0$.}
    \label{fig:filter_par}
\end{figure}

In the context of the SG models here, a similar parameter study can be performed to determine a suitable value for the filter parameter. Figure \ref{fig:filter_par} shows the expectation and standard deviation for a simple Burger's equation test case and different filter parameters $\alpha$. We choose a constant $s=3$ as the filter exponent $s$ is only modifying the shape of the filter strength in a mild way. Furthermore, we also choose $N_* = 0$ fixed as no additional variables need to remain unchanged. 
The results in figure \ref{fig:filter_par} clearly visualize that a small value of the filter parameter $\alpha$, e.g., $\alpha=1$, is not sufficient to damp the oscillations of both the expected values as well as the standard deviation. Similarly, a very large value of the filter parameter, e.g., $\alpha=60, 100$, also leads to oscillations. In between, there is a range of parameters, for which the oscillations become negligible. This includes the value $\alpha=36$, which was frequently used in the literature. This indicates that the choice of $\alpha=36$ also seems to perform well in the settings of this paper and we therefore use it in all test cases including the exponential filter.

\section*{Acknowledgments}
    This research has been partially supported by the European Union's Horizon 2020 research and innovation program under the Marie Sklodowska-Curie grant agreement no. 888596.
    Tianbai Xiao is funded by the Alexander von Humboldt Foundation (Ref3.5-CHN-1210132-HFST-P).
	Jonas Kusch is funded by the Deutsche Forschungsgemeinschaft (DFG, German Research Foundation) -- Project-ID 258734477 -- SFB 1173. Julian Koellermeier is a postdoctoral fellow in fundamental research of the Research Foundation -- Flanders (FWO), funded by FWO grant no. 0880.212.840. 

\bibliographystyle{unsrt}
\bibliography{main}

\end{document}